\documentclass[usenatbib,useAMS]{mn2e}

\usepackage{rotating}
\usepackage{graphicx}
\usepackage{epstopdf}
\usepackage{amsmath}
\usepackage{amssymb}
\usepackage{bbold}
\usepackage{multirow}
 
\usepackage{xcolor}

\usepackage{float}

\newcommand{\kms}{\mbox{km\,s$^{-1}$}}
\newcommand{\kuns}{\mbox{$h$\,Mpc$^{-1}$}}
\newcommand{\Muns}{\mbox{$h^{-1}\,{\rm Mpc}$}}

\newcommand{\mbi}[1]{\mbox{\boldmath$#1$}}

\newcommand{\mat}[1]{\mbox{\rm\bf #1}}
\newcommand{\lsim}{\mbox{${\,\hbox{\hbox{$ < $}\kern -0.8em \lower 1.0ex\hbox{$\sim$}}\,}$}}
\newcommand{\gsim}{\mbox{${\,\hbox{\hbox{$ > $}\kern -0.8em \lower 1.0ex\hbox{$\sim$}}\,}$}}

\def\etal{{\it et al.\ }}

\def\beqn{\vspace{2mm}
\begin{eqnarray}} 
\def\eeqn{\vspace{2mm} 
\end{eqnarray}}

 \def\la{\langle}

\newcommand{\be}{\begin{equation}}
\newcommand{\ee}{\end{equation}}
\newcommand{\ba}{\begin{eqnarray}}
\newcommand{\ea}{\end{eqnarray}}
\newcommand{\brr}{\begin{array}}
 
\newcommand{\err}{\end{array}}
\newcommand{\bc}{\begin{center}}
\newcommand{\ec}{\end{center}}

\begin{document}

\title[Cosmic velocity fields]{Estimating cosmic velocity fields from density fields and tidal tensors}

\author[Kitaura \etal]{Francisco-Shu Kitaura$^{1,2}$\thanks{E-mail: kitaura@aip.de, Karl-Schwarzschild fellow}, Raul E.~Angulo$^{2}$, Yehuda Hoffman$^{3}$  and Stefan Gottl\"ober$^{1}$\\
$^{1}$  Leibniz-Institut f\"ur Astrophysik (AIP), An der Sternwarte 16, D-14482 Potsdam, Germany \\
$^{2}$  Max-Planck Institut f\"ur Astrophysik (MPA), Karl-Schwarzschildstr.~1, D-85748 Garching, Germany \\
$^{3}$ Racah Institute of Physics, Hebrew University, Jerusalem, Israel 
}

\maketitle

\begin{abstract}
In this work we investigate the nonlinear and nonlocal relation between cosmological density and peculiar velocity fields. Our goal is to provide an algorithm for the reconstruction of the nonlinear velocity field from the fully nonlinear density.
We find that including the gravitational tidal field tensor using second order Lagrangian perturbation theory (2LPT) based upon an estimate of the linear component of the nonlinear density field significantly improves the estimate of the cosmic flow in comparison to linear theory not only in the low density, but also and more dramatically in the high density regions. In particular we test two estimates of the linear component: the lognormal model and the iterative Lagrangian linearisation. The present approach relies on a rigorous higher order Lagrangian perturbation theory analysis which incorporates a nonlocal relation. It does not require additional fitting from simulations being in this sense parameter free, it is independent of  statistical-geometrical optimisation and it is straightforward and efficient to compute. 
The method is demonstrated to yield an unbiased estimator of the  velocity field on scales  $\gsim$5 {\Muns} with closely Gaussian distributed errors. 
Moreover, the statistics of the divergence of the peculiar velocity field is extremely well recovered showing a good agreement with the true one from $N$-body simulations. The typical errors of about 10  {\kms} (1 sigma confidence intervals) are reduced by more than 80\%  with respect to linear theory in the scale range between 5 and 10 {\Muns} in high density regions ($\delta>2$). We also find that iterative Lagrangian linearisation is significantly superior in the low density regime with respect to the lognormal model. 
\end{abstract}



\begin{table*}
\rotatebox[]{0}{
\begin{tabular}{|c|c|c|c|} 
{\bf Works}& {\bf $\theta$-$\delta$ relation}&{\bf parameters}&{\bf methodology}\\\hline
{ \citet{1991ApJ...372..380Y}} & $\theta=\delta$ &$\{ \mbi p\}$& linear theory: LPT \\ 
  \citet{1991ApJ...379....6N}  & $\theta=\delta/(1+\alpha\delta)$ &$\alpha$, $\{ \mbi p\}$& empirical approximation\\ 
 \citet{vdB92}  & $\theta=\alpha\left[\left(1+\delta\right)^{\beta}-1\right]$  &$\alpha,\beta$, $\{ \mbi p\}$& PT \\
\citet{vdG93}  &  $\theta=\delta-\alpha D_1^2\mu^{(2)}(\phi_g)$&$\alpha, \{ \mbi p\}$& approx. 2LPT\\ 
\citet{vdW97}  & $\theta=\left[(1+\alpha^2\sigma^2_\delta)\delta+\alpha\sigma^2_\delta\right]/(1+\alpha\delta)$ &$\alpha,\sigma^2_\delta$, $\{ \mbi p\}$& empirical approximation\\ 
\citet{vdC98} &$\theta=\left[\delta-\alpha\left(\delta^2-\sigma^2_\delta\right)+\beta\delta^3\right]$&$\alpha,\beta,\sigma^2_\delta$, $\{ \mbi p\}$ &PT+$N$-body\\
\citet{vdB99} &$\theta=\left[\gamma\delta+\alpha\left(\delta^2-\sigma^2_\delta\right)+\beta\delta^3\right]$&$\alpha,\beta,\gamma,\sigma^2_\delta$, $\{ \mbi p\}$ &PT+$N$-body\\
 \citet{vdK00} & $\theta=\alpha\left[\left(1+\delta\right)^{1/\alpha}-1\right]+(\alpha-1)/(2\alpha)\sigma^2_\delta$ &$\alpha,\sigma^2_\delta$, $\{ \mbi p\}$ & PT+Eulerian grid-based code \\ 
 \citet{vdBC08} &  $\theta=\gamma\left[\left(1+\delta\right)^{1/\alpha}-\left(1+\delta\right)^{\beta}\right]$ &$\alpha,\beta,\gamma$, $\{ \mbi p\}$  & spherical collapse model   \\    
{ this work} &${\theta} = D_1\delta^{(1)}-D_2f_2/f_1\,\mu^{(2)}(\phi^{(1)})$&$\{ \mbi p\}$ & 2LPT\\    
\end{tabular}
}
\caption{ \label{tab:vdrel} The parameters $\alpha,\beta,\gamma$ are either
derived from first principles \citep[][this work]{vdB92,vdG93,vdBC08} or from fitting to simulations
being different for each case. These parameters also depend on the scale of
interest.  The parameters which are in parenthesis are fully determined by the chosen
cosmology and the theoretical model. The variance of the density field is given
by $\sigma_\delta^2=\langle\delta^2\rangle$. PT stands for perturbation theory
and is applied only in the univariate case (local relation). 2LPT stands for
second order Lagrangian perturbation theory and yields the only nonlocal (and
nonlinear) relation from the list. \citet{1995A&A...298..643H,1999MNRAS.308..763M} proposed 2LPT to correct for redshift distortions.  Let us mention here the least-action
principle methods to determine the peculiar velocities from mass tracer objects
(galaxies) introduced by \citet[][]{Peebles89} and further extended by
\citet[][]{NB00,BEN02,LMCTBS08}.  }
\end{table*}

\section{introduction}

Gravitational instability is one of the key ingredients to explain the rich
hierarchy of structures we observe today in the Universe.  It has amplified
small mass fluctuations produced after inflation to give rise from large galaxy
clusters to huge voids. Simultaneously, the same local gravitational field
imprinted ``peculiar'' velocities in galaxies; deviations from the overall
expansion of the Universe which is responsible for the Hubble flow.

The peculiar velocity of galaxies is a valuable quantity in cosmology since it
contains similar but complementary information to that enclosed in the galaxies
position.  For instance, by requiring isotropy in the measured galaxy
clustering, the cosmological mass density parameter and even the nature of
gravity can be constrained \citep[see e.g.][]{1996ApJ...473...22D,
1998ApJ...507...64W, 2001MNRAS.326.1191B,2008Natur.451..541G}.  In addition,
these motions can be used to reconstruct the properties of the universe at an
earlier time, in principle, even at recombination where perturbations were
completely linear \citep[][]{1992ApJ...391..443N,
1993ApJ...405..449G,1997MNRAS.285..793C,1998ApJ...508..440N,1999MNRAS.308..763M,2002Natur.417..260F}.

A method able to accurately determine the peculiar velocity field can be used
in many different applications; ranging from bias studies combining galaxy
redshift surveys with measured peculiar velocities \citep[see
e.g.][]{1995MNRAS.272..885F,1999ApJ...520..413Z,2011arXiv1109.3856C}, Baryon
Acoustic Oscillations reconstructions \citep[see e.g.][]{2007ApJ...664..675E},
determination of the growth factor, to estimates of the initial conditions of
the Universe which in turn can be used for constrained simulations \citep[see
e.g.][]{2010arXiv1005.2687G,2003ApJ...596...19K}. A particularly well suited
application regards the topological methods to detect the kinematic
Sunyaev-Zeldovich effect.

There is in addition recent interest in the measurement of large-scale flows.
Some authors claim to have detected a so-called ``dark flow'' caused by
super-horizont perturbations \citep[see
e.g.][]{2009ApJ...691.1479K,2011ApJ...732....1K}. Others discuss such flows as
a challenge to the standard cosmological model as a whole \citep[see
e.g.][]{2009MNRAS.392..743W}.

Unfortunately, the apparent shift in spectral features of a galaxy is also
affected by the expansion of the universe, therefore it is not possible to
directly measure the peculiar velocity field in spectroscopic surveys. For this
reason, one has to resort to indirectly infer it from the mass density
fluctuations \citep[but see][for a recent alternative
method]{2011arXiv1106.6145N}.  However, this is not a trivial procedure due the
highly nonlinear state of the density fluctuations today and due to its
nonlocal relationship with the velocity field. 

The simplest approach is given by the linear continuity equation, which is
routinely used in clustering studies. However, it has a range of applicability
only limited to very large scales \citep[e.g.][]{Angulo2008}. Alternative
methods devised to improve upon linear theory can be separated into three
areas. The first one consists on developing nonlinear relationships with
higher-order perturbation theory \citep[][]{vdB92,vdC98,vdB99,vdK00}, with
spherical collapse models \citep[][]{vdBC08} or based  on empirical relations
found in cosmological N-body simulations \citep[][]{1991ApJ...379....6N,vdW97}.

Another idea is to solve the boundary problem of finding the initial positions
of a set of particles governed by the Eulerian equation of motion and gravity
with the least action principle \citep[see][]{Peebles89,NB00,BEN02}.  A similar
approach consists on relating the observed positions of matter tracers (e.g.
galaxies) in a geometrical way to a homogeneous distribution by minimizing a
cost function, which combined with the Zeldovich approximation
\citep{1970A&A.....5...84Z} provides an estimation of the velocity field
\citep[see][]{2005ApJ...635L.113M,LMCTBS08}.

 The diversity of strategies and approximations for obtaining the velocity from
the density field  hint at the difficulty of the problem.  Some approaches are
simply not accurate enough and some are computationally very expensive. This
sets the agenda for an improvement in the field. Any new method should be accurate,
unbiased, computationally efficient and applicable to observational data.

 A further shortcoming of the existing approaches is that they are mostly
particle-based, which is not applicable for more general matter tracers like the
Lyman-alpha forest or the 21-cm line, nor they can be combined with optimal
density field estimators \citep[see][]{kitaura_log,jasche_hamil,kitaura_lyman}.

In this paper we investigate a different approach based on higher order Lagrangian perturbation
theory, and it is motivated by the pioneering work of \citet{vdG93} and further extended
by \citet{1995A&A...298..643H,1999MNRAS.308..763M}.  The theoretical basis for
LPT was carefully worked out by
\citet[][]{1991ApJ...382..377M,1993MNRAS.264..375B,1994MNRAS.267..811B,bouchet1995,1995MNRAS.276..115C}
\citep[for further references see][]{2002PhR...367....1B}.

Contrary to classic applications of LPT, in which the properties of an evolved
distribution are predicted from a linear density field in Lagrangian
coordinates (e.g. in the generation of initial conditions for $N$-body
simulations or of galaxy mock catalogues:
\citet{2010MNRAS.403.1859J,2002MNRAS.329..629S}), our starting point is an
evolved field in Eulerian coordinates (e.g. the present-day galaxy
distribution). The key realisation of our approach is that it is possible to
decompose a nonlinear density field into a Gaussian and Non-Gaussian term,
which are related to each other through LPT \citep[see similar
approaches][]{1993ApJ...405..449G,vdG93,1999MNRAS.308..763M}. In other words,
it is possible to find a closely Gaussian field which would evolve, under the
assumption of LPT, into the measured density field today. This Gaussian field
can then naturally be used to predict the corresponding velocity field today in
LPT.


 Our method combines i) the equations of motion and continuity for a fluid
under self gravity in 2-nd (3-rd) order Lagrangian Perturbation theory (LPT)
with ii) the idea that the present-day galaxy distribution can be expanded into
a closely linear-Gaussian field and a highly skewed nonlinear component
consistent with 2LPT (or 3LPT) \citep[see][]{kitaura_gauss}. The former aspect
makes our approach physically motivated and also captures the nonlocal nature
of the density-velocity relation via the gravitational tidal field tensor. The
latter aspect self-consistently minimises the impact of the approximations of a
2nd order formulation of gravitational evolution, but more importantly, it
enables the application of LPT to nonlinear fields.  

We note that the use of the lognormal transformation (including the subtraction of a mean
field) to obtain an estimate of the linear field was proposed by
\citet[][]{kitaura_lyman} and has been recently confirmed to give already a
good estimate of the divergence of the displacement field
\citep[][]{2011arXiv1111.4466F}.  To estimate the velocity field one needs,
however, to go to higher order perturbation theory as we show in this work.

In the next section we recap Lagrangian perturbation theory and derive
the velocity-density relation to second and third order. In section \ref{sec:method}
we will present the method to compute the peculiar velocity field from the
nonlinear density field. In section \ref{sec:tests} we will present our
numerical tests based on the Millennium Run simulation. Here we will analyze
the goodness of the recovered velocity divergence as compared to the true one
and the same for each velocity component. A study of the errors in our method
is also presented. Finally we present our conclusions.

\section{Velocity--density relation}
\label{sec:theory}

The first part of this section presents the relation between density and
velocity fields in 2LPT, and how it can be applied to an evolved field. In
the second part, we outline a practical implementation of this method.

\subsection{Second order Lagrangian perturbation theory}

The basic idea in Lagrangian perturbation theory is that an initial homogeneous
field expressed in Lagrangian coordinates $\mbi q$ can be related to a final
field in Eulerian coordinates $\mbi x$ trough a unique mapping: $\mbi x=\mbi
q+\mbi\Psi(\mbi q)$ determined by a displacement field $\mbi\Psi$ \citep[see e.g.][]{2002PhR...367....1B}. 

The linear solution for this expression is the well-known Zeldovich
approximation, in which the displacement field is given by the Laplacian of the
gravitational potential at $\mbi q$. This result has been successfully applied to many
aspects of cosmology, but it fails to describe the dynamics of a nonlinear field
where shell crossings and curved trajectories are common. 

An improvement is found by expanding the displacement field and considering 
higher order terms \citep[see e.g.][]{1994A&A...288..349B,1995A&A...294..345M,bouchet1995}. For instance, 
the displacement field to second order is given by 

 \be 
\label{eq:2lpt} 
 \mbi \Psi (\mbi q)  =  - D_1\nabla_{q}\phi^{(1)}(\mbi q) + D_2\nabla_{q}\phi^{(2)}(\mbi q),
\ee

\noindent where 
$D_1$ is the linear growth factor normalised to unity today, $D_2$ the
second order growth factor, which is given by $D_2 \simeq-3/7\,
\Omega^{-1/143}D_1^2$ \citep[see][]{1993MNRAS.264..375B,bouchet1995}.  The potentials $\phi^{(1)}(\mbi q)$ and
$\phi^{(2)}(\mbi q)$ are obtained by solving a pair of Poisson equations:
$\nabla^2\phi^{(1)}(\mbi q) = \delta^{(1)}(\mbi q)$, where $\delta^{(1)}$ is the linear
over-density, and $\nabla^2\phi^{(2)}(\mbi q) =  \delta^{(2)}(\mbi q)$.

It is important to realise that these terms are not independent of each other.
The second order nonlinear term $\delta^{(2)}$ is fully determined by the 
linear over-density field $\delta^{(1)}$ through the
following quadratic expression

\be
\label{eq:2lptdelta2}
 \delta^{(2)}(\mbi q) \equiv\mu^{(2)}(\phi^{(1)}(\mbi q) )=\sum_{i>j} 
    \Big( \phi^{(1)}(\mbi q) _{,ii}\phi^{(1)}(\mbi q) _{,jj}-
    [\phi^{(1)}(\mbi q) _{,ij}]^2\Big)\,,
\ee
where we use  the following notation
$\phi_{,ij}(\mbi q)  \equiv \partial^2\phi(\mbi q) /\partial q_i\partial q_j$, and the indices $i,j$
run over the three Cartesian coordinates.

Similarly, the particle co-moving velocities ${\mbi v}$ are given to second
order by:

\be
\label{eq:2lptvel}
 {\mbi v} (\mbi q) =-D_1f_1H\nabla_q\phi^{(1)}(\mbi q)+D_2f_2H\nabla_q\phi^{(2)}(\mbi q)\,, 
\ee

\noindent where  $f_i = {\rm d}\ln D_i/{\rm d}\ln a$, $H$ is the Hubble
constant and $a$ is the scale factor.  For flat models with a non-zero
cosmological constant, the following relations apply $f_1\approx\Omega^{5/9},
f_2\approx2\Omega^{6/11}$ \citep[see][]{bouchet1995}, where $\Omega(z)$ is the
matter density at a redshift $z$.

We note that going to third order in Lagrangian perturbation theory provides
modest improvements
\citep[see][]{1993MNRAS.264..375B,1995A&A...294..345M,1995MNRAS.276..115C,bouchet1995,2002PhR...367....1B}.

\begin{figure*}
\begin{tabular}{cc}
\includegraphics[width=8.5cm]{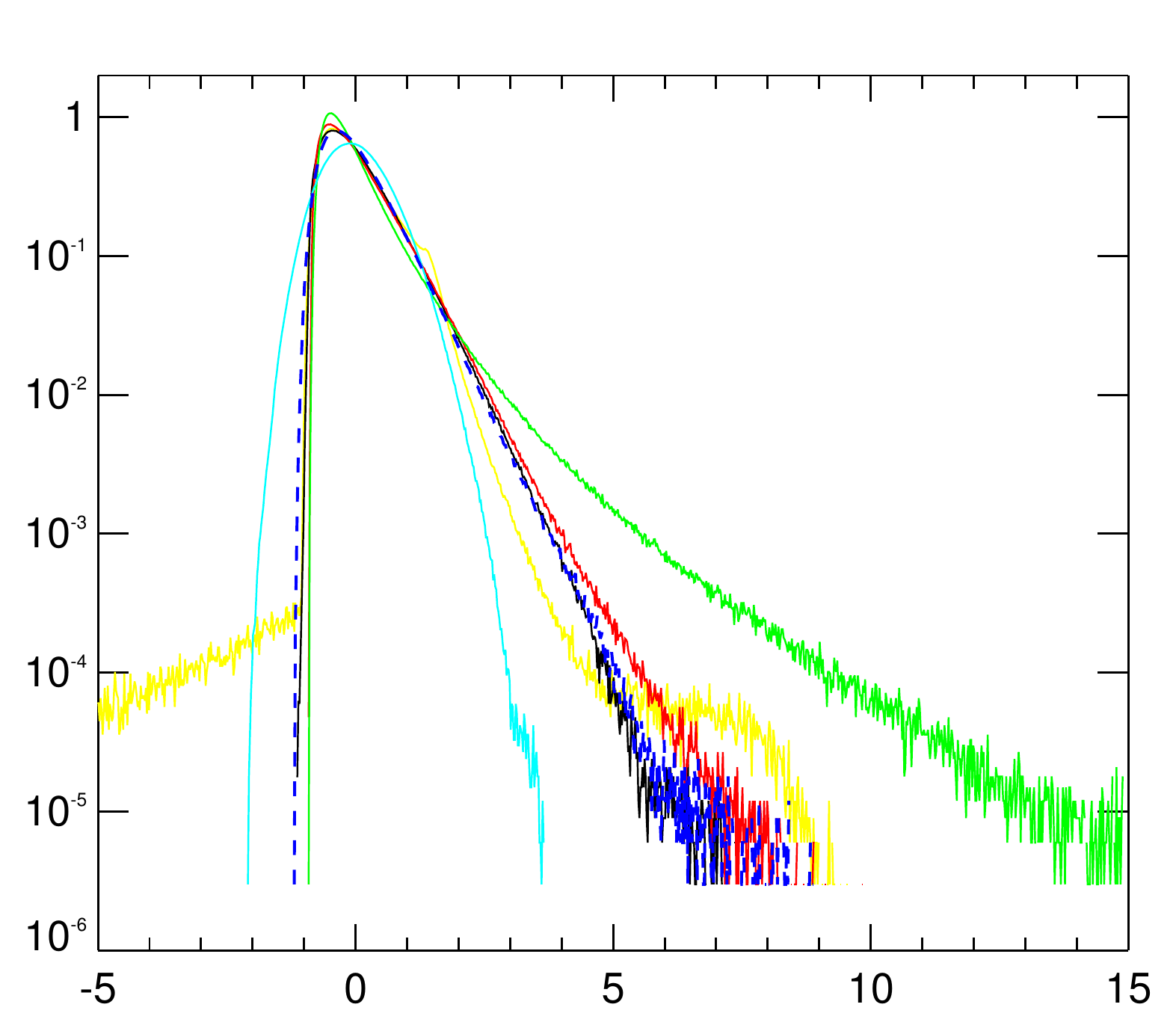}
\put(-85,165){{\large $r_{\rm S}$=5 {\Muns}}}
\put(-250,105){\rotatebox[]{90}{{PDF}}}
\put(-120,-10){{$\theta_k$}}
\put(-70,130){\color{blue}Nbody}
\put(-70,100){\color{green}LIN}
\put(-70,150){\color{yellow}GRAM}
\put(-70,140){\color{cyan}LOG}
\put(-70,120){\color{black}LOG-2LPT}
\put(-70,110){\color{red}2LPT}
\includegraphics[width=8.5cm]{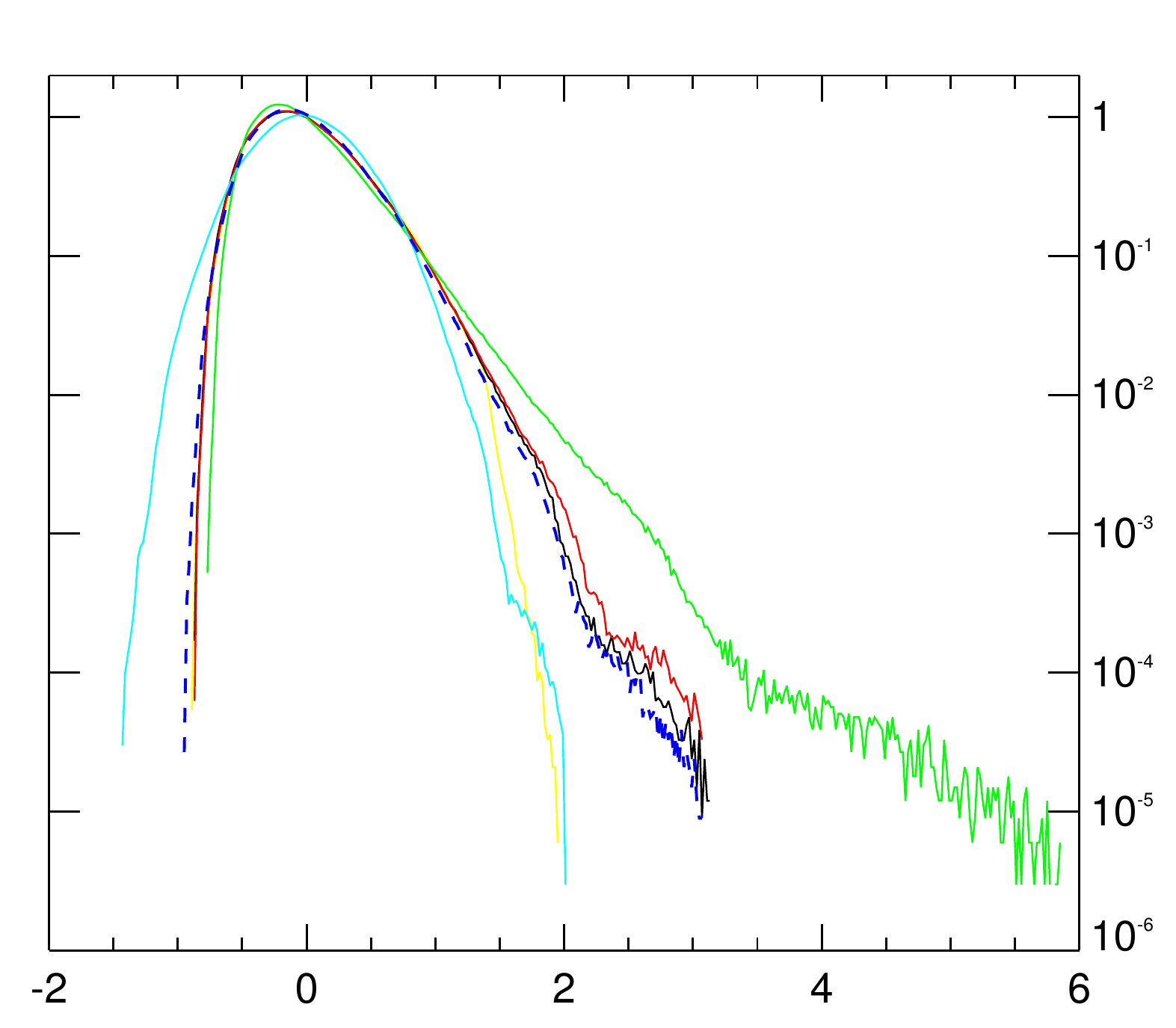}
\put(-105,165){{\large $r_{\rm S}$=10 {\Muns}}}
\put(-120,-10){{$\theta_k$}}
\end{tabular}
\caption{\label{fig:stats2}  
 Statistics of the scaled velocity divergence ($\theta_k$: true from the
simulation $\theta^{\rm Nbody}$ and different reconstructed ones $\theta^{\rm
rec}$ as explained below) with different smoothing scales: Left: $r_{\rm S}=$ 5
{\Muns}, Right: 10 {\Muns}. Curves from left to right in the order of
appearance: yellow: approximate 2LPT solution from the nonlinear density field
(GRAM) (see G93 in table \ref{tab:vdrel}), cyan: logarithm of the density field
(LOG), blue dashed: true scaled velocity divergence at $z=0$ the Millennium Run
(Nbody), black: 2LPT estimate from the logarithm of the density field
(LOG-2LPT) (Eq.~\ref{eq:2lpttheta}), red:  2LPT estimate from the iterative
solution (2LPT) (see \S\ref{sec:method}), green: linear theory (LIN) (density
field). } 
\end{figure*}
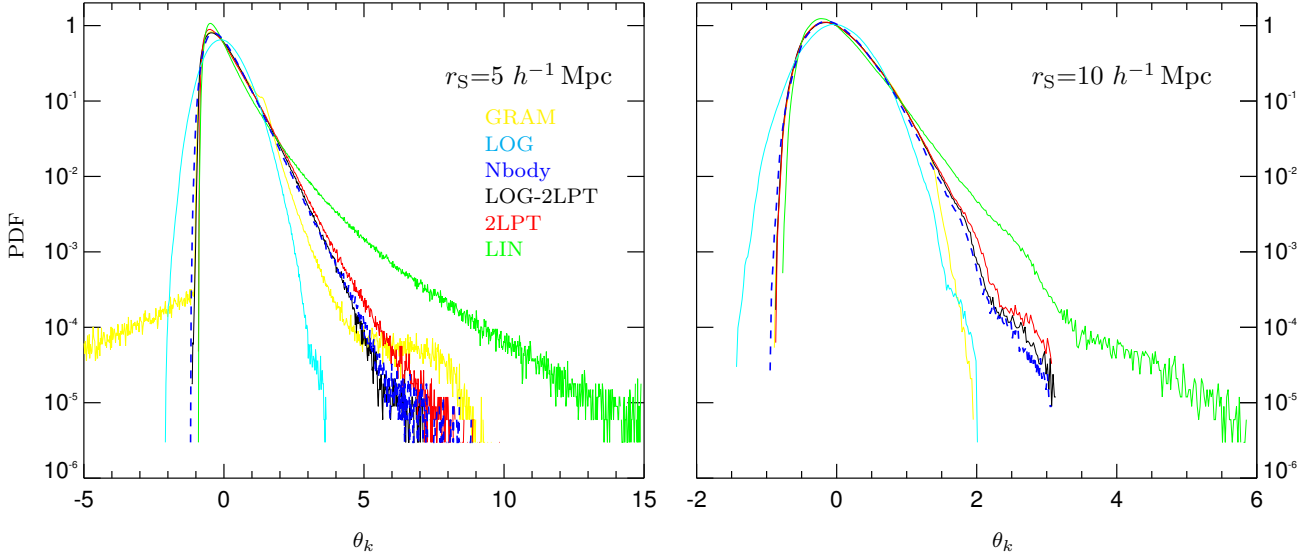

{\color{black}





 To apply the Lagrangian framework to a density field in the Eulerian frame $\delta_{\rm M}(\mbi x)$ one must be careful. Under the assumption that there is no shell-crossing, one can write the inverse equation relating Eulerian to Lagrangian coordinates $\mbi q=\mbi
x-\mbi\Psi(\mbi x)$. Mass conservation leads to the following equation \citep[see][]{1991ApJ...379....6N}:

\be
\label{eq:jacE}
1+\delta_{\rm M}(\mbi x)= \tilde{\mat J}(\mbi x) \,,
\ee
with
\be
 \tilde{\mat J}(\mbi x) \equiv \left|\frac{\partial\mbi q}{\partial\mbi x}\right|\,.
\ee

Expanding the Jacobian $ \tilde{\mat J}(\mbi x)$ one finds \citep{kitaura_gauss};

\ba
\label{eq:2lptsplit0}
  \delta_{\rm M}(\mbi x)&=&\left|1-\nabla_{x}\cdot\mbi \Psi(\mbi x)\right|-1\nonumber\\
&\simeq&-\nabla_x\cdot\mbi \Psi(\mbi x)+\mu^{(2)}(\Theta(\mbi x))+\mu^{(3)}(\Theta(\mbi x))\,.
\ea
The displacement or velocity field derived from this expression will automatically be expressed in Eulerian coordinates as we will show below.
We note that the same expression is found as a function of the Lagrangian coordinate $\mbi q$ when expanding the inverse of the corresponding Jacobian  $\mat J(\mbi q)\equiv\left|{\partial\mbi x}/{\partial\mbi q}\right|$  \citep[see][]{kitaura_gauss}.

Using the displacement field (Eq.~\ref{eq:2lpt}) in Eulerian coordinates, the final density field can be written in terms of the linear and nonlinear
fields:

\ba
\label{eq:2lptsplit}
\lefteqn{\delta_{\rm M}(\mbi x)}\nonumber\\
&&\hspace{-0.5cm}=D_1\delta^{(1)}(\mbi x)-D_2\mu^{(2)}(\phi^{(1)}(\mbi x))+\mu^{(2)}(\Theta(\mbi x))+\mu^{(3)}(\Theta(\mbi x))\nonumber\\
&&\hspace{-0.5cm}=\delta^{\rm L}(\mbi x)+\delta^{\rm NL}(\mbi x)\,,
\ea

\noindent {\color{black} with $\Theta(\mbi x)$ being the potential associated to the divergence of the displacement field: $\Theta(\mbi x)=-\nabla^{-2}\nabla\cdot\mbi \Psi(\mbi x)$}, 
$\delta^{\rm L}(\mbi x)=D_1\delta^{(1)}(\mbi x)$ and $\delta^{\rm NL}(\mbi x)=-D_2\mu^{(2)}(\phi^{(1)}(\mbi x))+\mu^{(2)}(\Theta(\mbi x))+\mu^{(3)}(\Theta(\mbi x))$. The third order term in the Jacobian expansion is given by:
\be
\mu^{(3)}(\Theta(\mbi x))=\det\left(\Theta(\mbi x)_{,ij}\right)\,.
\ee 
 From now on the Eulerian coordinate dependence $(\mbi x)$ is left out. 
}

Assuming that any primordial vorticity has no growing modes associated (the
first vorticity terms appear in 3rd order Lagrangian perturbation theory, see
appendix) implies that the velocity field today is fully characterised by its
divergence ($\nabla\cdot\mbi v$), or, for convenience, by the scaled velocity
divergence, defined as:

\be
\theta\equiv-{f}(\Omega,\Lambda,z)^{-1}\nabla\cdot\mbi v\,,
\ee

\noindent with ${f}(\Omega,\Lambda,z)\equiv\dot{D}_1/D_1=f_1(\Omega,\Lambda,z)\,H(z)$. 

Combining Eqs.~\ref{eq:2lptsplit} and \ref{eq:2lptvel} truncated to quadratic terms in $\Phi^{(1)}$: $\delta^{\rm NL}=\left(D_1^2-D_2\right)\delta^{(2)}$,
one gets
\be
\label{eq:2lpttheta3}
 {\theta} =\delta_{\rm M}-\left[D_1^2+\left(\frac{f_2}{f_1}-1\right)D_2 \right]\delta^{(2)}\,. 
\ee

This expression is very similar to the one found by \citet{vdG93} (see
Tab.~\ref{tab:vdrel}).  Note, however, that using directly the evolved field as
the source for the second order term $\delta^{(2)}$ is a good approximation for
the true velocity field only on very large scales (where $\delta_{\rm M}$ is
close to unity), as we will see in \S\ref{sec:tests} and Fig.~\ref{fig:stats2},
breaking down on scales of even 10 {\Muns} for both estimations of the
nonlinear field and of the velocity divergence.

In this paper we follow a different approach, and solve iteratively the following
equation;

\be
\label{eq:2lpttheta}
 {\theta} = D_1\delta^{(1)}-D_2\frac{f_2}{f_1}\delta^{(2)}\,. 
\ee

\noindent which results from taking the divergence of Eq.~\ref{eq:2lptvel}. For this, we rely
on an estimation of the linear component of the present-day density field,
which in turn can be calculated by solving iteratively
Eq.~(\ref{eq:2lptsplit}). 
{\color{black} Note that the Eulerian description forces one to expand the inverse of the Jacobian \citep[see Eqs.~\ref{eq:jacE}-\ref{eq:2lptsplit} and][]{kitaura_gauss}. This is the main difference with respect to the work  of \citet[][]{1999MNRAS.308..763M} in which first the Jacobian is expanded and then the inverse of it is taken and could explain why 
we avoid  problems caused by artificial Lagrangian caustics in low density regions as reported in their work.}

 In practice, a
good approximation for the linear term, $\delta^{(1)}$, is simply given by the logarithm of the
density field; $\delta^{\rm L}=D_1\delta^{(1)}=\ln(1+\delta_{\rm M})-\mu$, with
$\mu=\langle\ln(1+\delta_{\rm M})\rangle$, as shown by
\citep{2009ApJ...698L..90N,kitaura_gauss}. Note that this expression is
essentially the lognormal approximation for the matter field
\citep[][]{1991MNRAS.248....1C}.  This local transformation has the advantage
of being computationally inexpensive.

In summary, approach finds a linear field which when plugged
into 2LPT expressions produces the observed matter field (or third order, see
appendix). If gravity worked only at a second order level, then this linear
field would be identical to the actual linear field that originated
$\delta_{\rm M}$, but naturally in reality it is just an approximation. Thus, 
it is important to characterise the performance and accuracy of the method, which 
we do in \S3. But first, we discuss a practical implementation of our method
in the next subsection. 

\subsection{Method} \label{sec:method}

The method to estimate the peculiar velocity field from the nonlinear density
field is straightforward and fast to compute. We now outline the steps to be
followed for its implementation. For this, we have assumed that there is an
unbiased and complete estimation of the matter field $\delta_{\rm M}$.  The
extra layer of complication introduced by shot noise, a survey mask, biasing
and redshift space distortions is out of the scope of this paper and will be
addressed in a future publication.

  \begin{enumerate}

\item  {\bf Linear density field}

One starts by computing an estimate of the linear component of the density field.
We propose two alternatives for this:

  \begin{enumerate}
  \item Lognormal model

\be
 \delta^{\rm L}_{\rm LOG}=D_1\delta^{(1)}=\ln(1+\delta_{\rm M})-\mu
\ee
  \item Gaussianisation with LPT \citep[][]{kitaura_gauss}

\be
 \delta^{\rm L}_{\rm LPT}=D_1\delta^{(1)}=\delta_{\rm M}- \delta^{\rm NL}
\ee
  \end{enumerate}

  \item {\bf Linear potential}

Then the  Poisson equation is solved to obtain the linear potential:
\be
\phi^{(1)} =  \nabla^{-2}\delta^{(1)}
\ee
 
 \item {\bf  Nonlinear second order density field}

The tidal field contribution to second order is calculated from the linear
potential:

\be
 \delta^{(2)}\left[\phi^{(1)}\right]=\sum_{i>j} 
    \Big( \phi^{(1)}_{,ii}\phi^{(1)}_{,jj}-
    (\phi^{(1)}_{,ij})^2\Big),
\ee

  \item {\bf Scaled velocity divergence}

One can now construct the 2nd order divergence of the velocity field taking
both the linear and the second order contribution:

\be  
 {\theta} = D_1\delta^{(1)}-D_2\frac{f_2}{f_1}\delta^{(2)}, 
\ee

  \item {\bf Peculiar velocity field}

Finally, one obtains the 3D velocity field:
\be
\mbi v=-{f}(\Omega,\Lambda,z)\nabla\nabla^{-2}\theta\,.
\ee
  \end{enumerate}

 Please note that the equations in steps (ib), (ii), (iii) and (v) can be
solved with FFTs. Details of the Gaussianisation step with LPT can be found in
\citet[][]{kitaura_gauss}. 

\section{Testing the method with $N$-body simulations}
\label{sec:tests}

In this section we test the performance of the method outlined above by
comparing the velocity field directly extracted from a $N$-body simulation with
our estimation based on the respective nonlinear density field. 

With this purpose, we employ the Millennium simulation which tracks the
nonlinear evolution of more than 10 billion particles in a box of comoving
side-length $500$ {\Muns} \citep{2005Natur.435..629S}. In particular, we
use the output corresponding to redshift $0$, which is where the most nonlinear
structures are present.

At such output we first compute the velocity and density field by mapping the
information of dark matter particles onto a grid of $256^3$ cells using the
nearest-grid-point (NGP) assignment scheme, which gives a spatial resolution of
about $2\,${\Muns}. We then apply the algorithm presented in \S2.2 to infer
the velocity divergence on a grid of identical dimensions. In the next two
subsections we present our results and explore the accuracy when applied on two
scales; 10 {\Muns} on which linear theory is usually assumed to perform
well, and 5 {\Muns} which enters into the mildly nonlinear regime.

\subsection{The velocity divergence}

\begin{figure*}
\begin{tabular}{cc}
\includegraphics[width=8.cm]{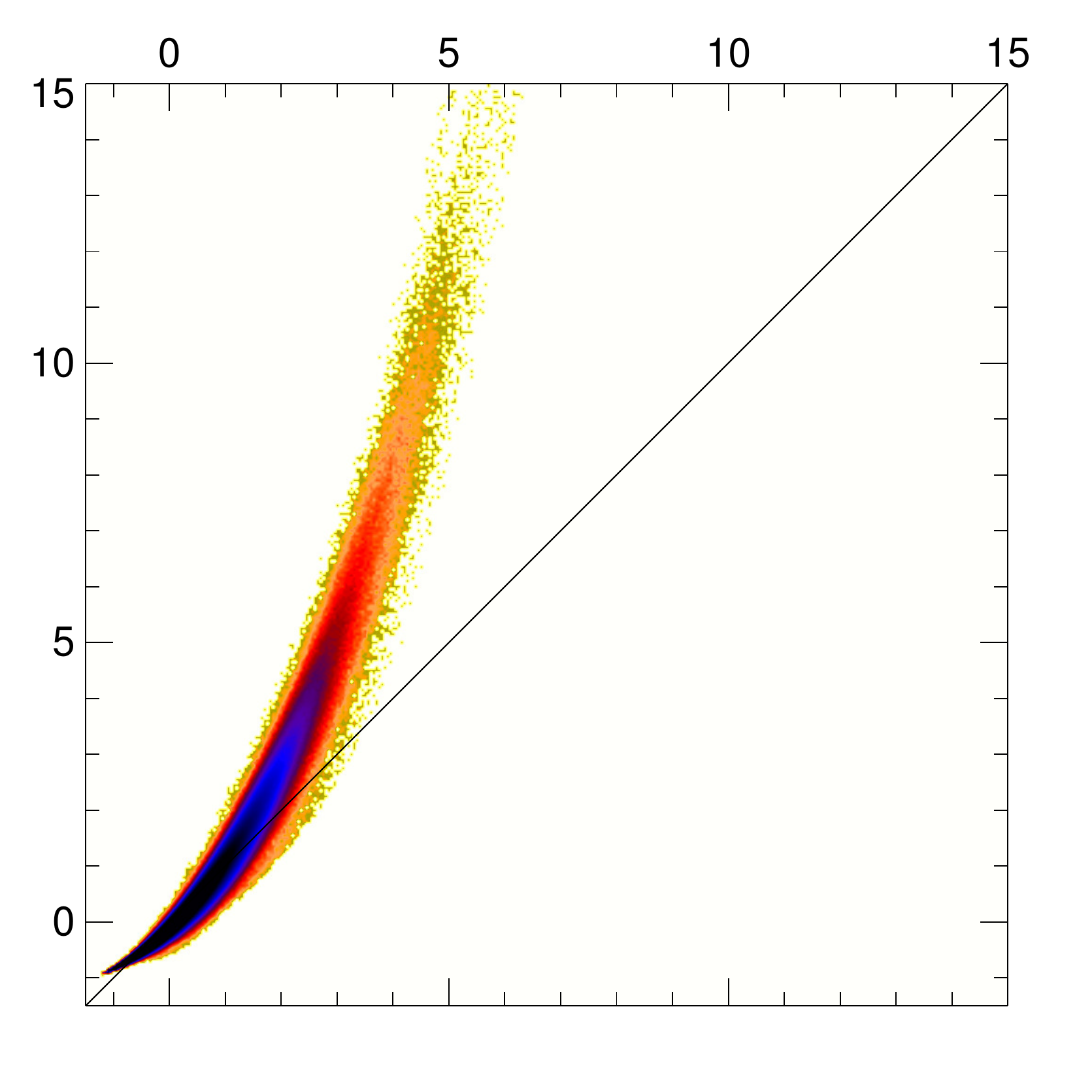}
\put(-100,70){{\large $r_{\rm S}$=5 {\Muns}}}
\put(-100,50){{\large LIN}}
\put(-230,110){\rotatebox[]{90}{{$\theta^{\rm rec}_{\rm LIN}$}}}
\hspace{-1.0cm}
\includegraphics[width=8.cm]{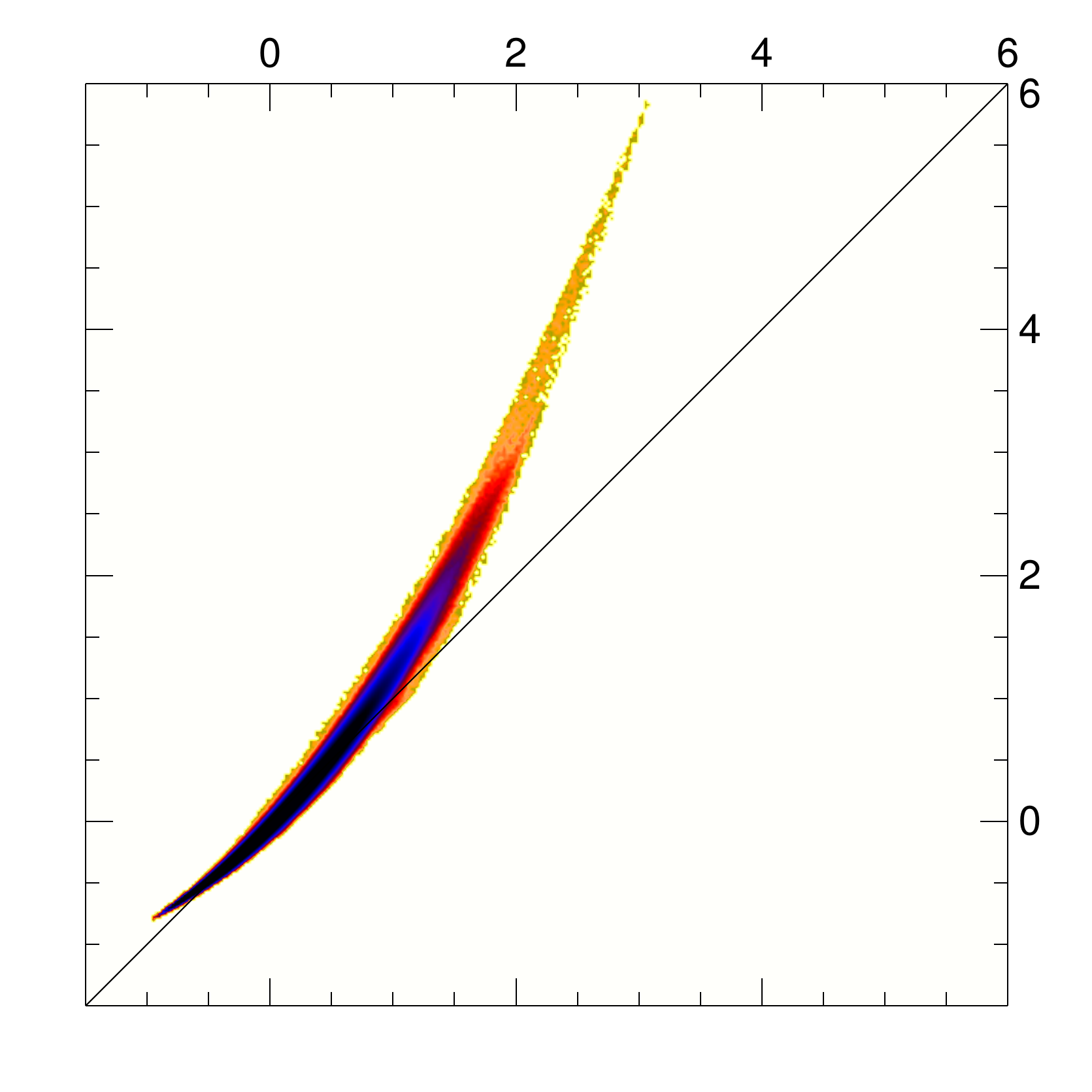}
\put(-100,70){{\large $r_{\rm S}$=10 {\Muns}}}
\put(-100,50){{\large LIN}}
\vspace{-1.0cm}
\\
\includegraphics[width=8.cm]{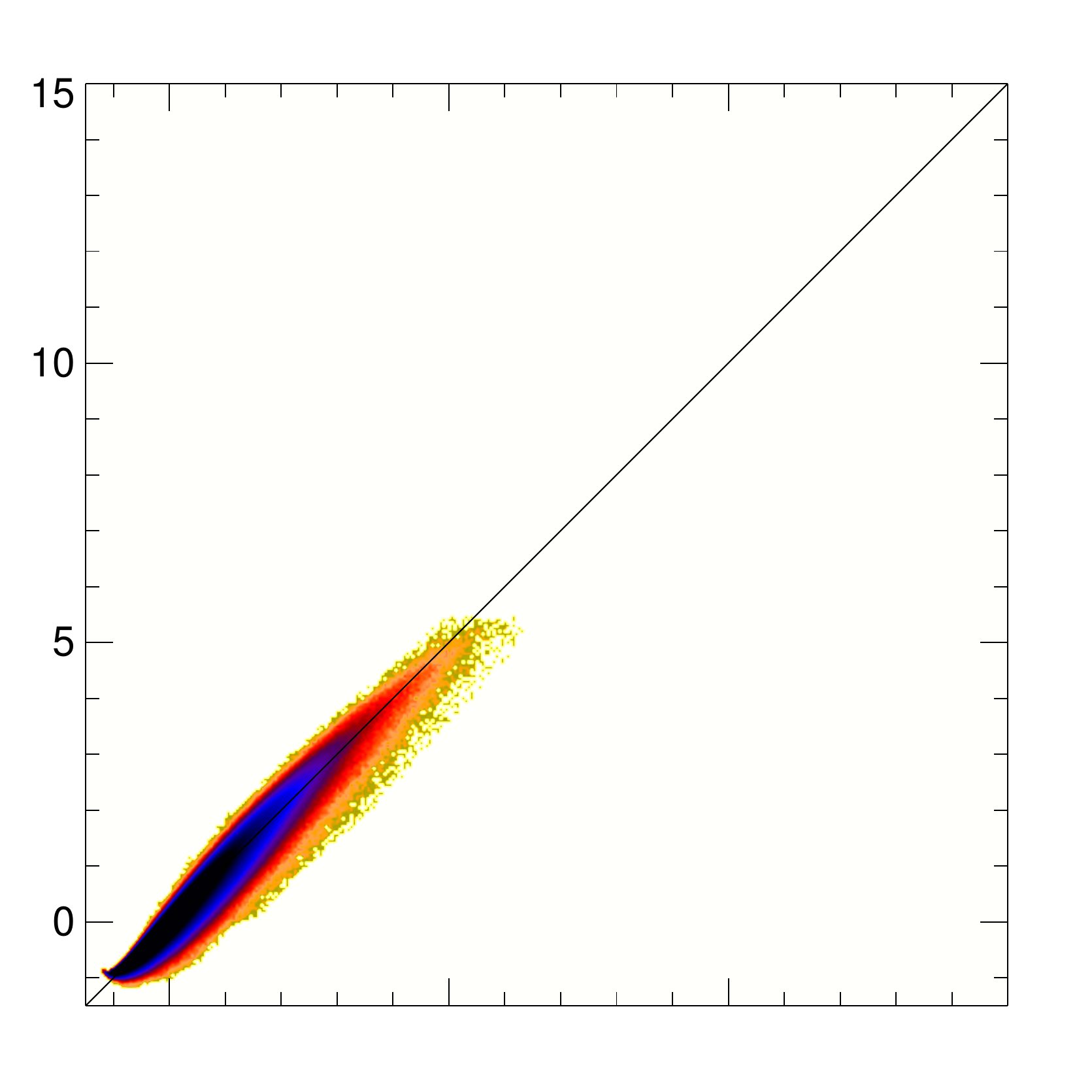}
\put(-230,110){\rotatebox[]{90}{{$\theta^{\rm rec}_{\rm LOG-2LPT}$}}}
\put(-100,70){{\large $r_{\rm S}$=5 {\Muns}}}
\put(-100,50){{\large LOG-2LPT}}
\hspace{-1.0cm}
\includegraphics[width=8.cm]{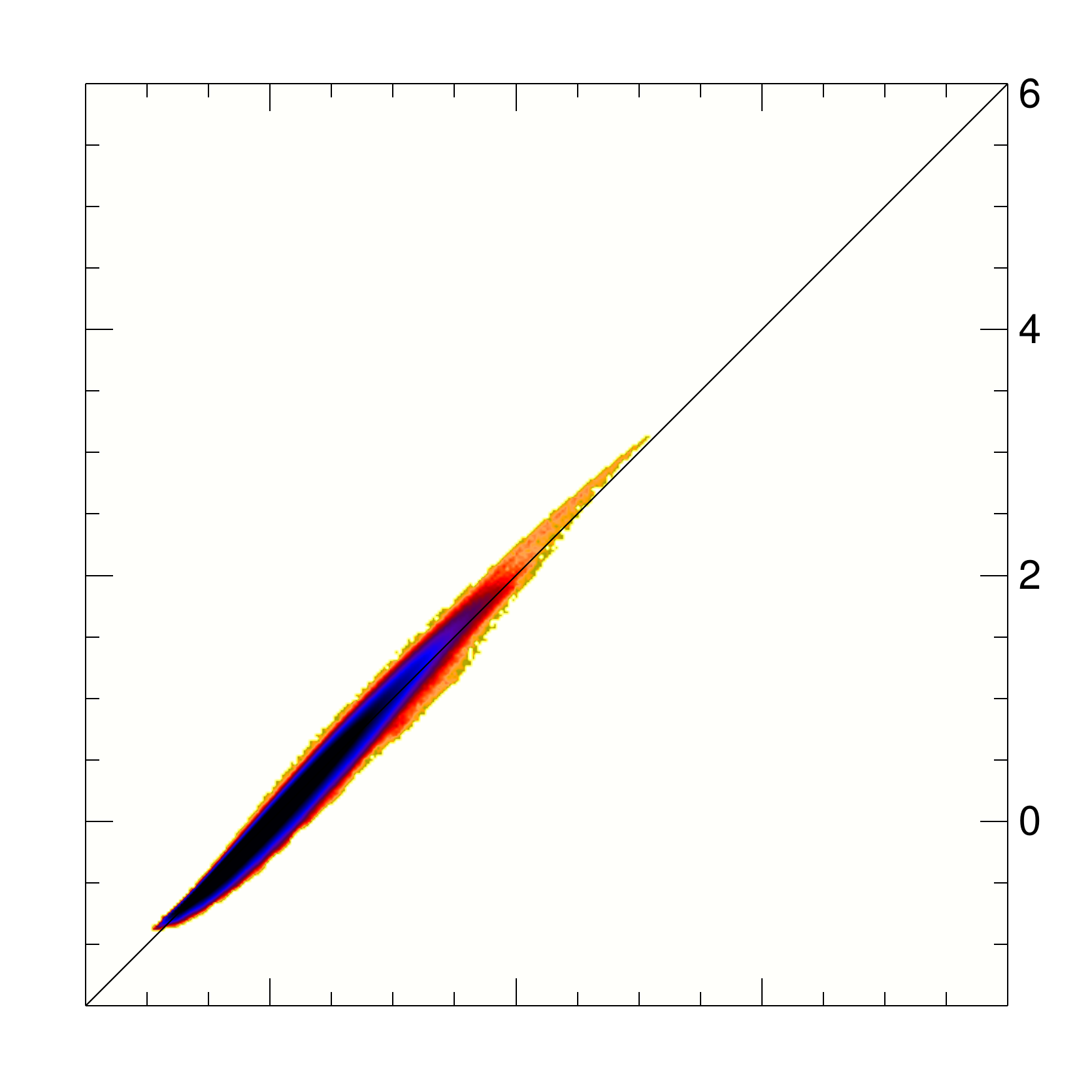}
\put(-100,70){{\large $r_{\rm S}$=10 {\Muns}}}
\put(-100,50){{\large LOG-2LPT}}
\vspace{-1.0cm}
\\
\includegraphics[width=8.cm]{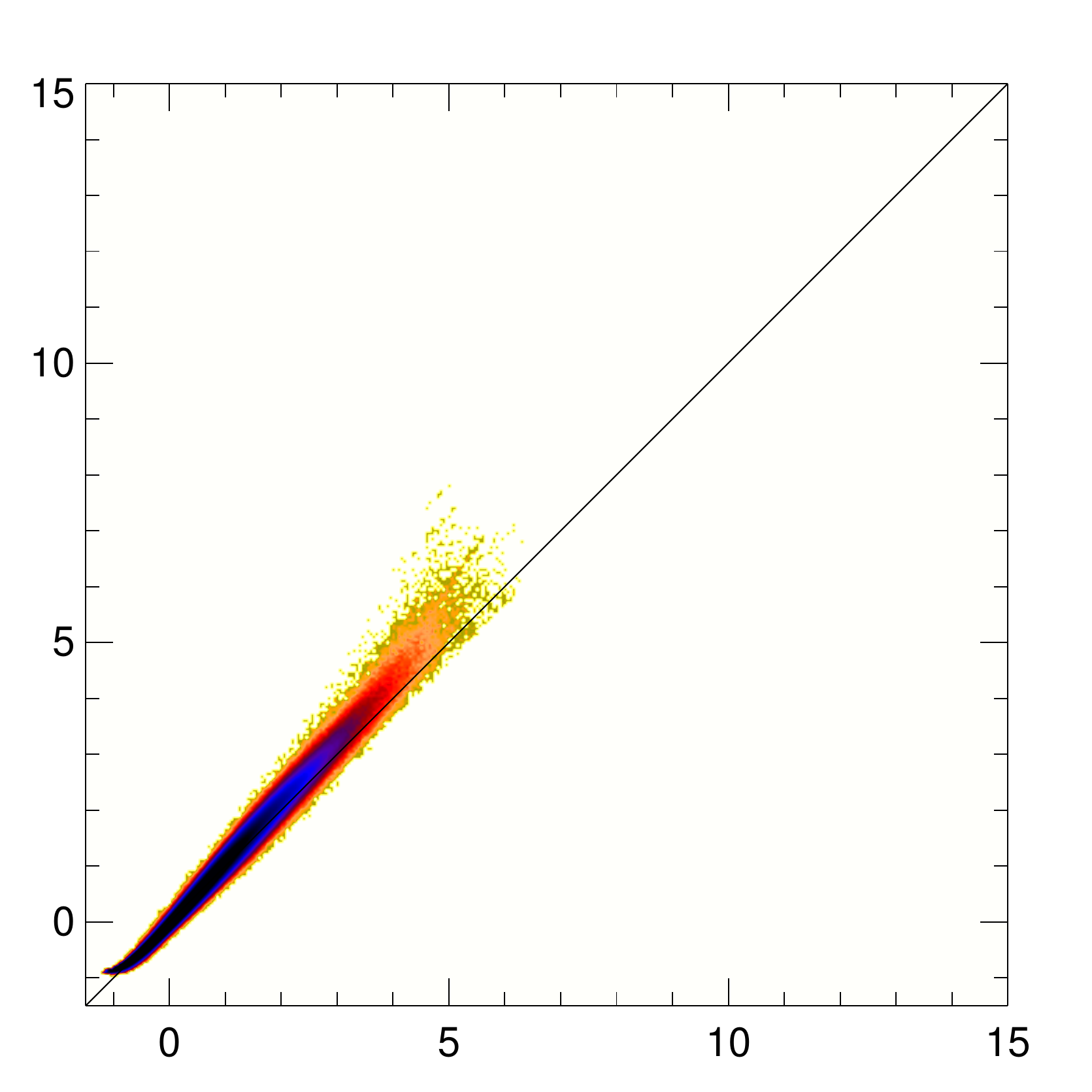}
\put(-230,110){\rotatebox[]{90}{{$\theta^{\rm rec}_{\rm 2LPT}$}}}
\put(-100,70){{\large $r_{\rm S}$=5 {\Muns}}}
\put(-100,50){{\large 2LPT}}
\put(-110,-5){{$\theta^{\rm Nbody}$}}
\hspace{-1.0cm}
\includegraphics[width=8.cm]{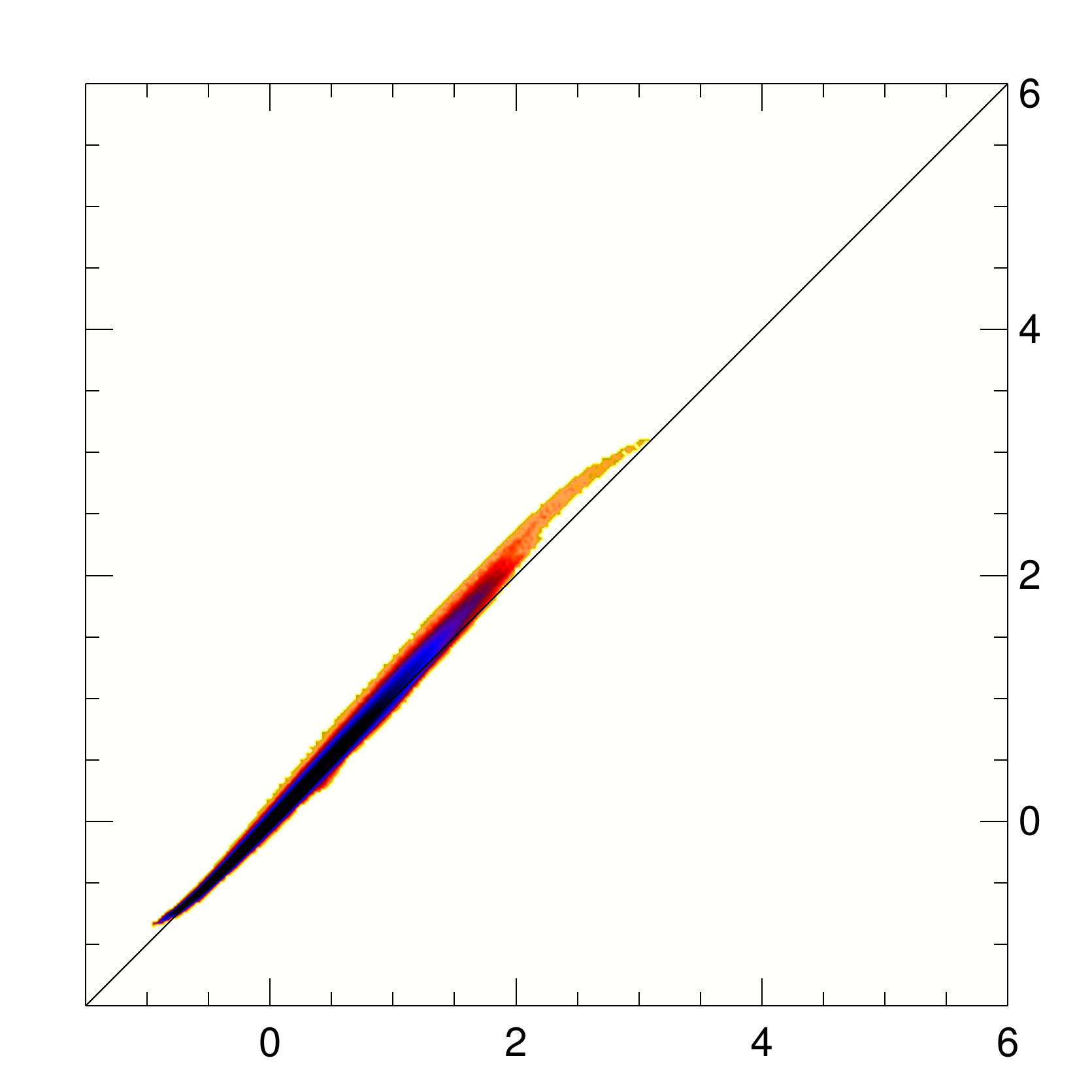}
\put(-100,70){{\large $r_{\rm S}$=10 {\Muns}}}
\put(-100,50){{\large 2LPT}}
\put(-110,-5){{$\theta^{\rm Nbody}$}}
\vspace{0.cm}
\end{tabular}
\caption{\label{fig:thetadiff} Cell-to-cell comparison between the {\it true} and the reconstructed scaled velocity divergence, $\theta^{\rm Nbody}$ and $\theta^{\rm rec}$ respectively. Left panels show results on scales of 5 {\Muns} and right panels on scales of 10 {\Muns}. Upper panels: LIN, middle panels: LOG-2LPT, lower panels: 2LPT. {\color{black} The dark colour-code indicates a high number and the light colour-code a low number of cells. }}
\end{figure*}

\begin{figure*}
\begin{tabular}{cc}
\includegraphics[width=7.8cm]{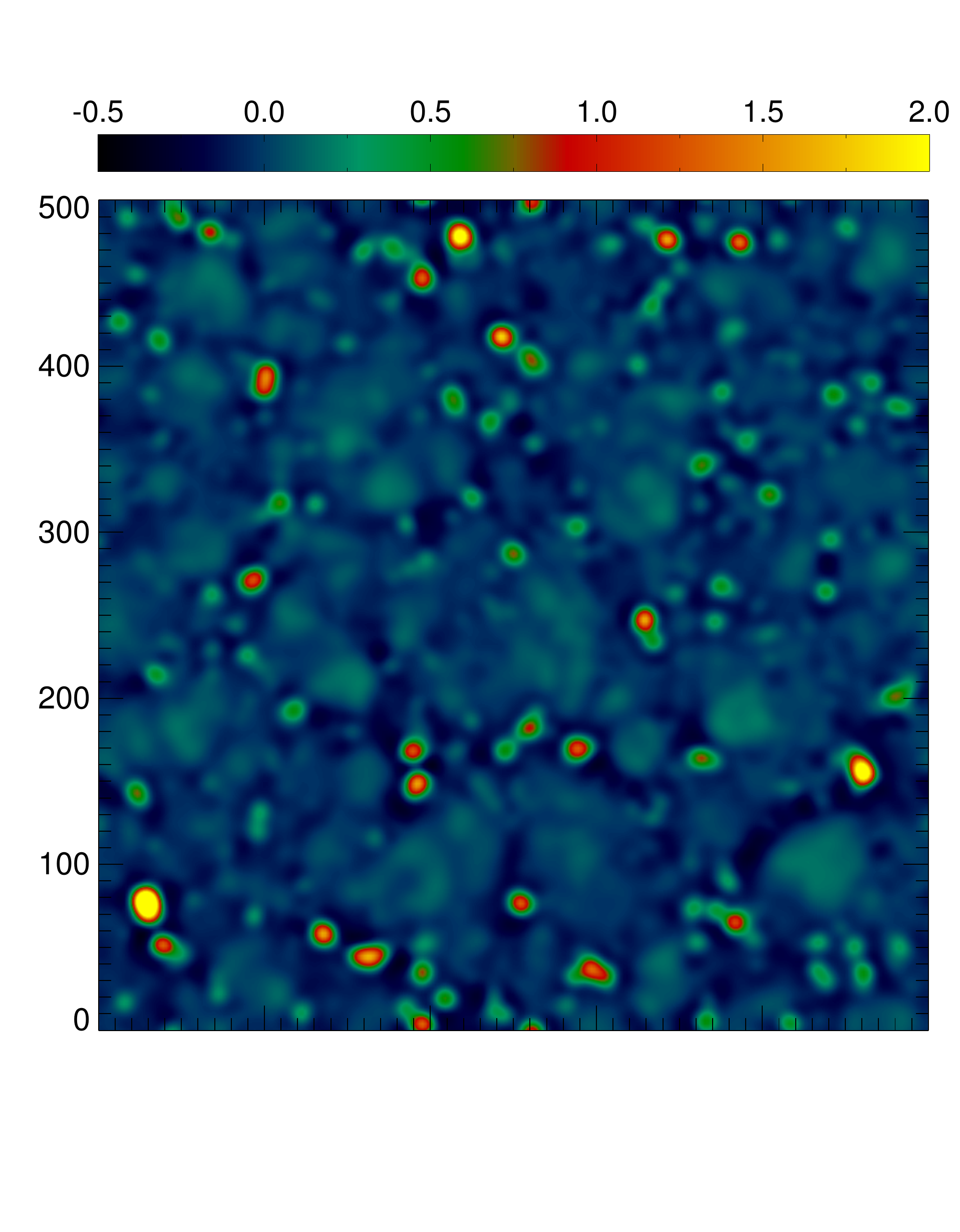}
\put(-225,140){\rotatebox[]{90}{{$Y$ [{\Muns}]}}}
\hspace{-.7cm}
\includegraphics[width=7.8cm]{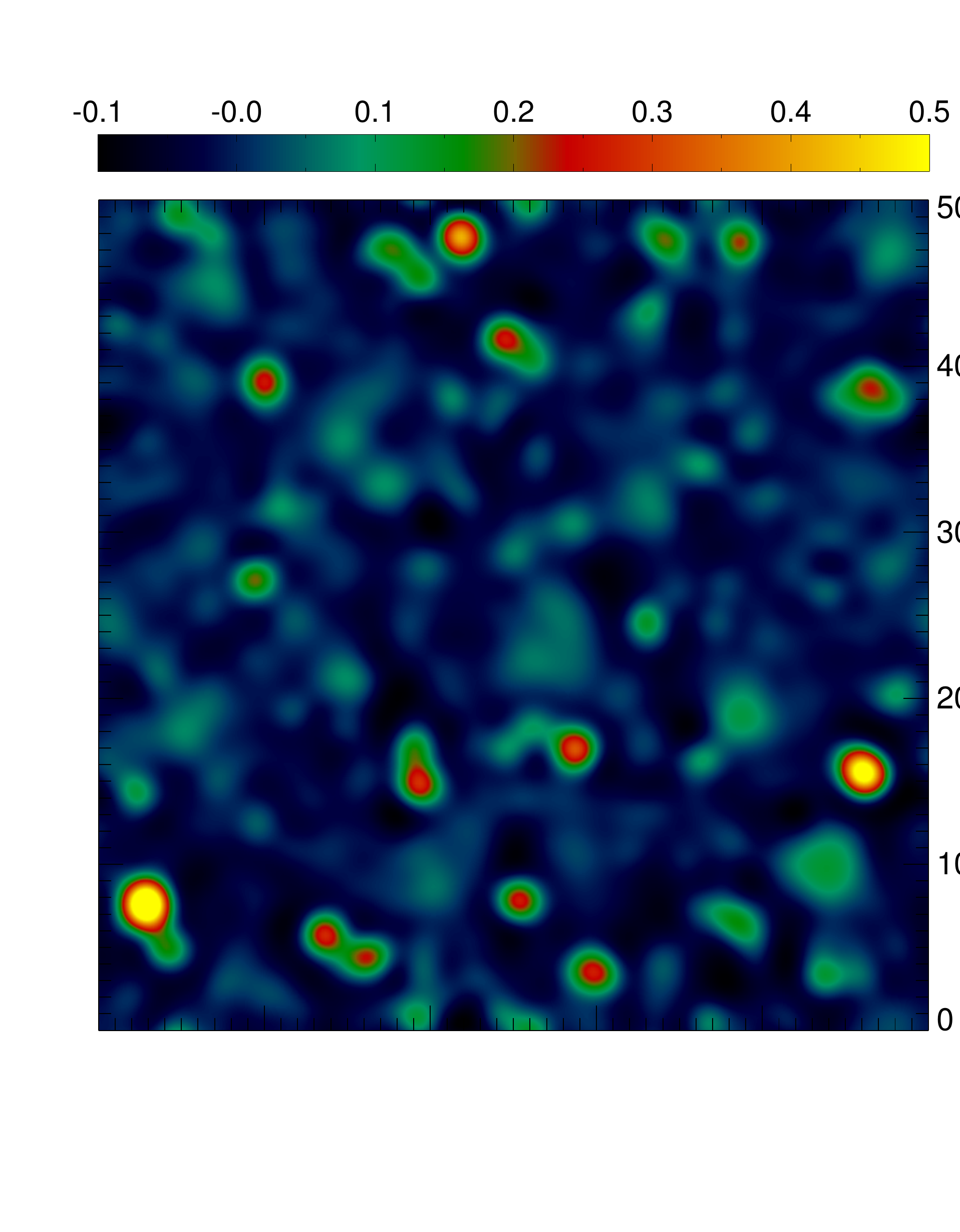}
\vspace{-2.9cm}
\\
\includegraphics[width=7.8cm]{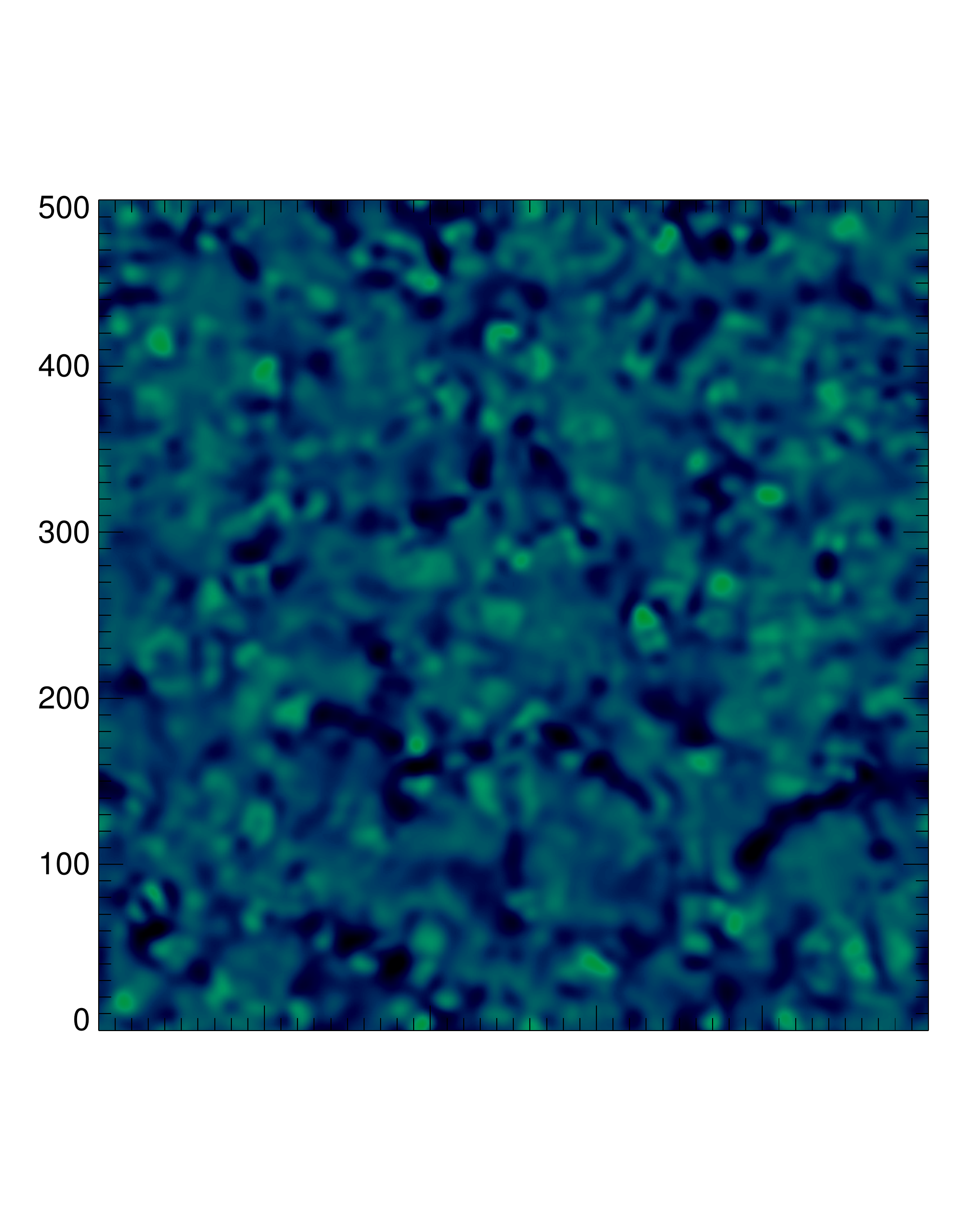}
\put(-225,140){\rotatebox[]{90}{{$Y$ [{\Muns}]}}}
\hspace{-.7cm}
\includegraphics[width=7.8cm]{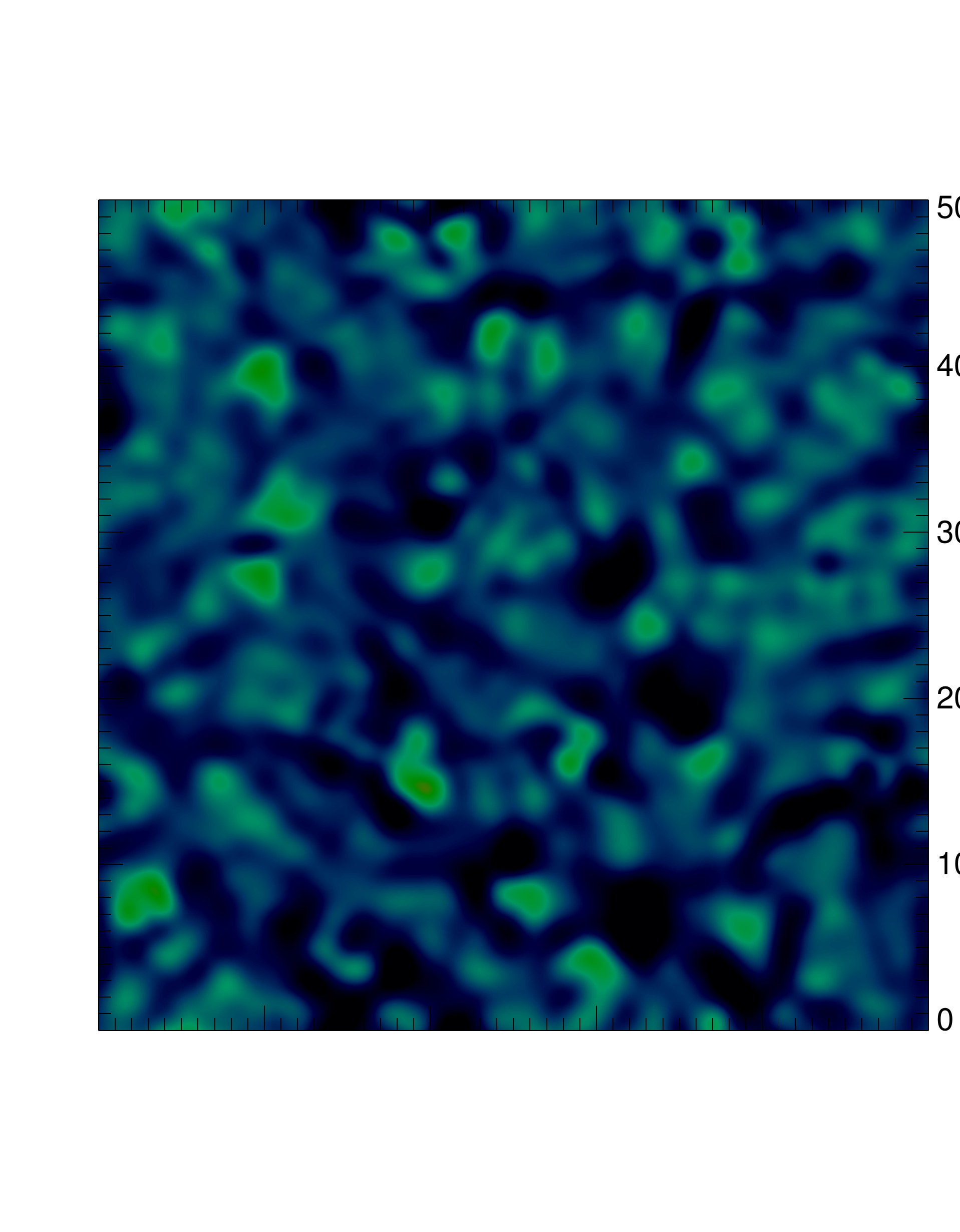}
\vspace{-2.9cm}
\\
\includegraphics[width=7.8cm]{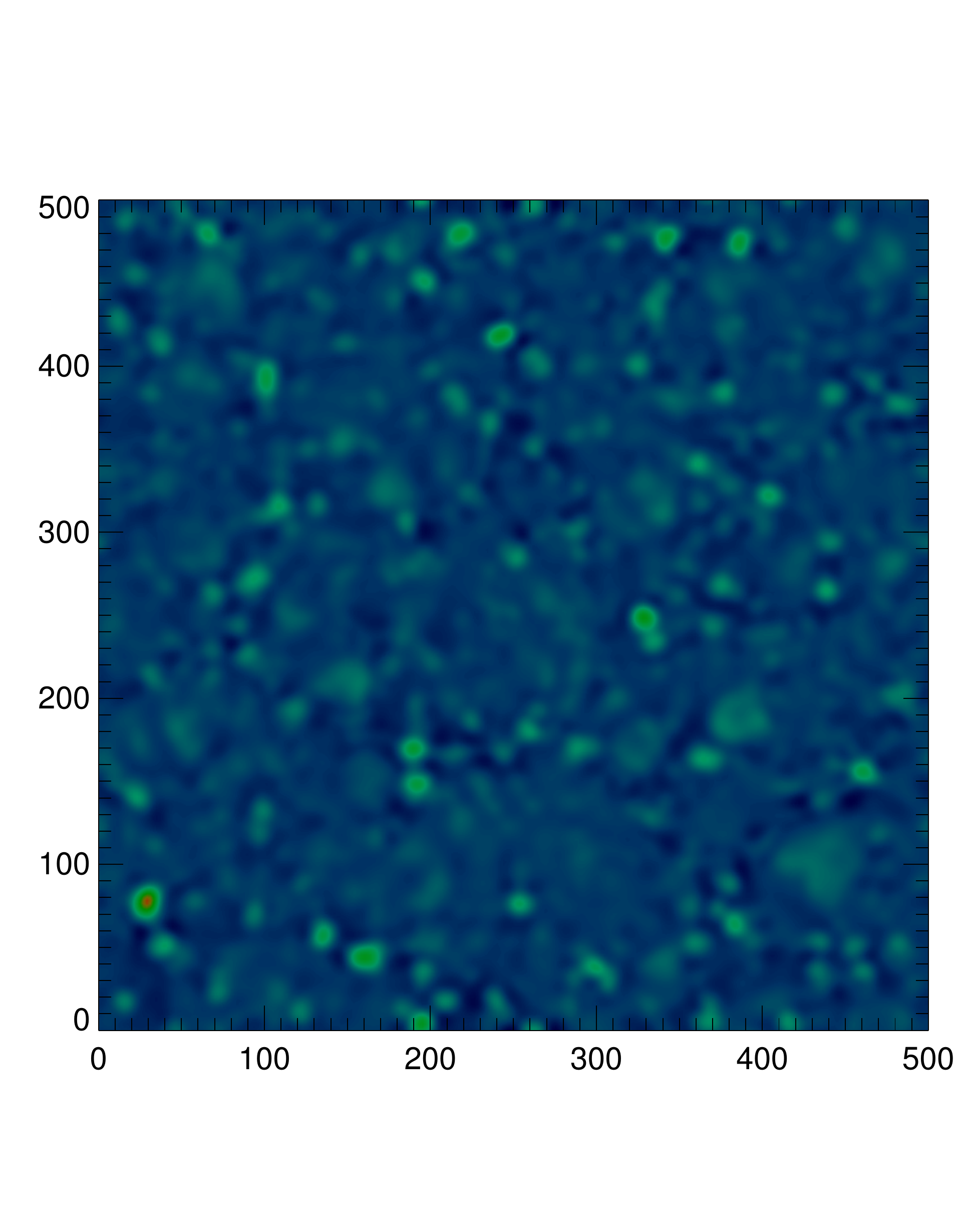}
\put(-225,140){\rotatebox[]{90}{{$Y$ [{\Muns}]}}}
\put(-130,25){{$X$ [{\Muns}]}}
\hspace{-.7cm}
\includegraphics[width=7.8cm]{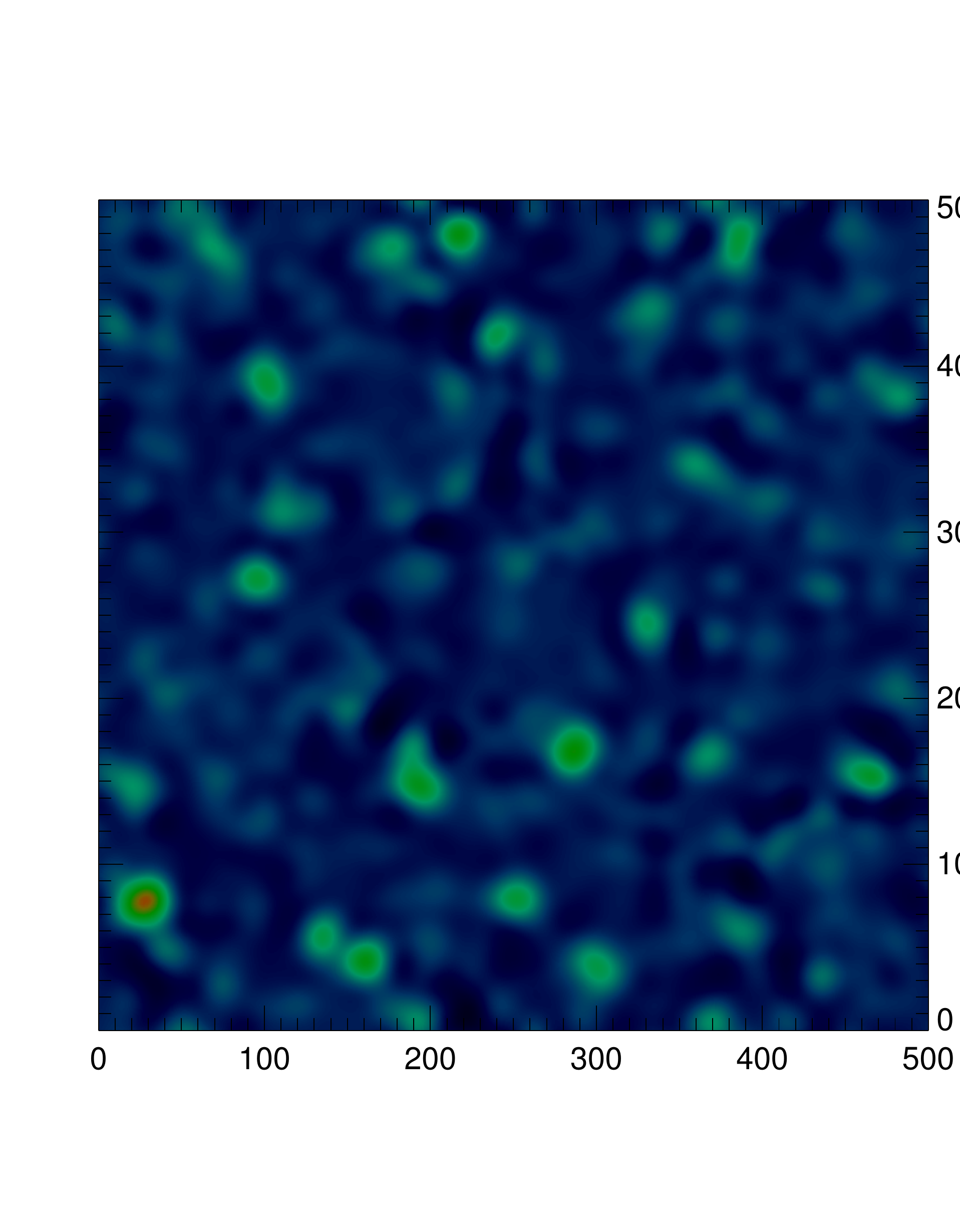}
\put(-130,25){{$X$ [{\Muns}]}}
\vspace{-1.cm}
\end{tabular}
\caption{\label{fig:corrtheta} Difference between the true and the estimated  scaled velocity divergence. On scales of $r_{\rm S}=$5 {\Muns} (left panels) and 10 {\Muns} (right panels) with different colour ranges. Upper panels: LIN, middle panels:  LOG-2LPT, lower panels: 2LPT.}
\end{figure*}

\begin{figure*}
\begin{tabular}{cc}
\includegraphics[width=7.8cm]{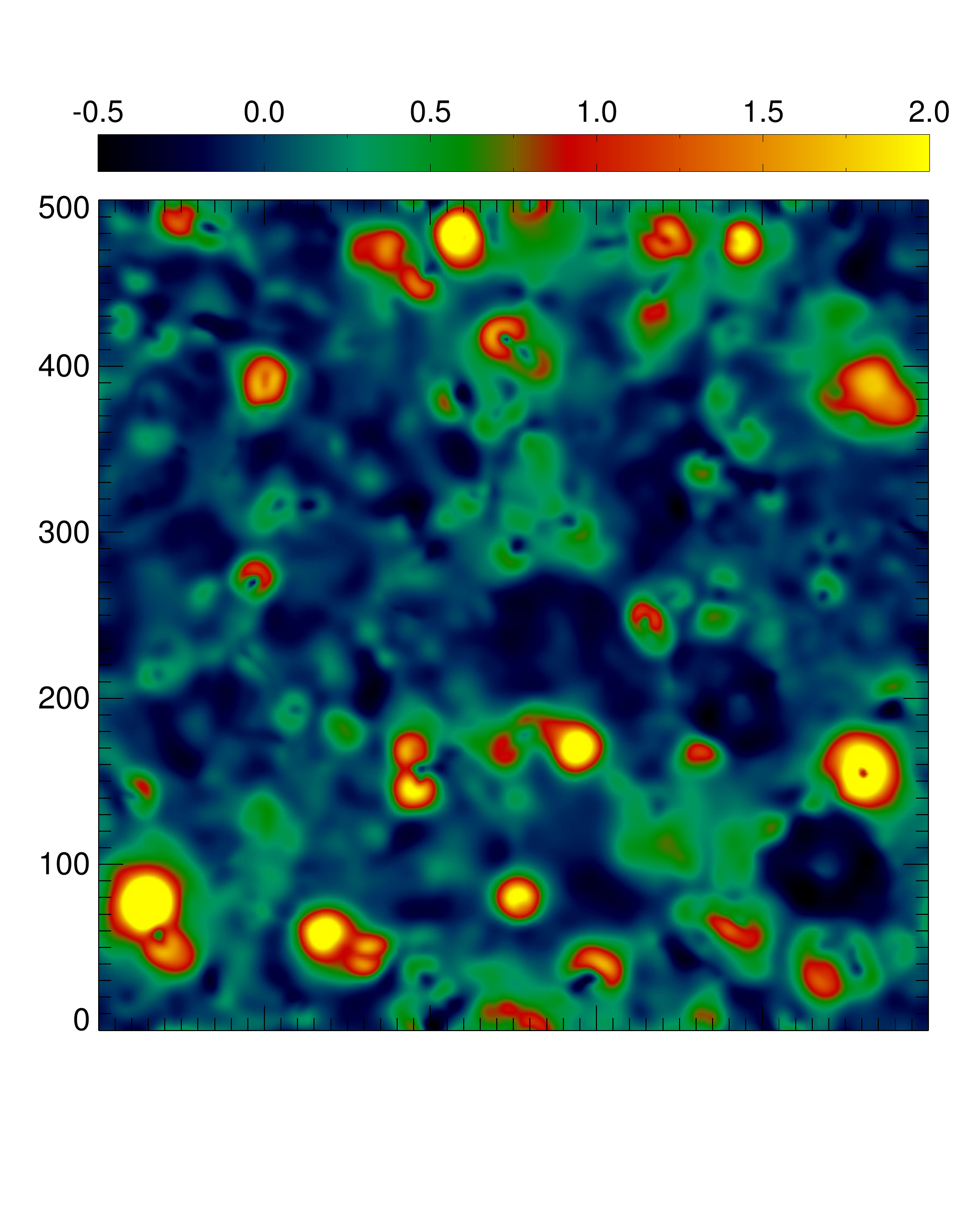}
\put(-225,140){\rotatebox[]{90}{{$Y$ [{\Muns}]}}}
\hspace{-.7cm}
\includegraphics[width=7.8cm]{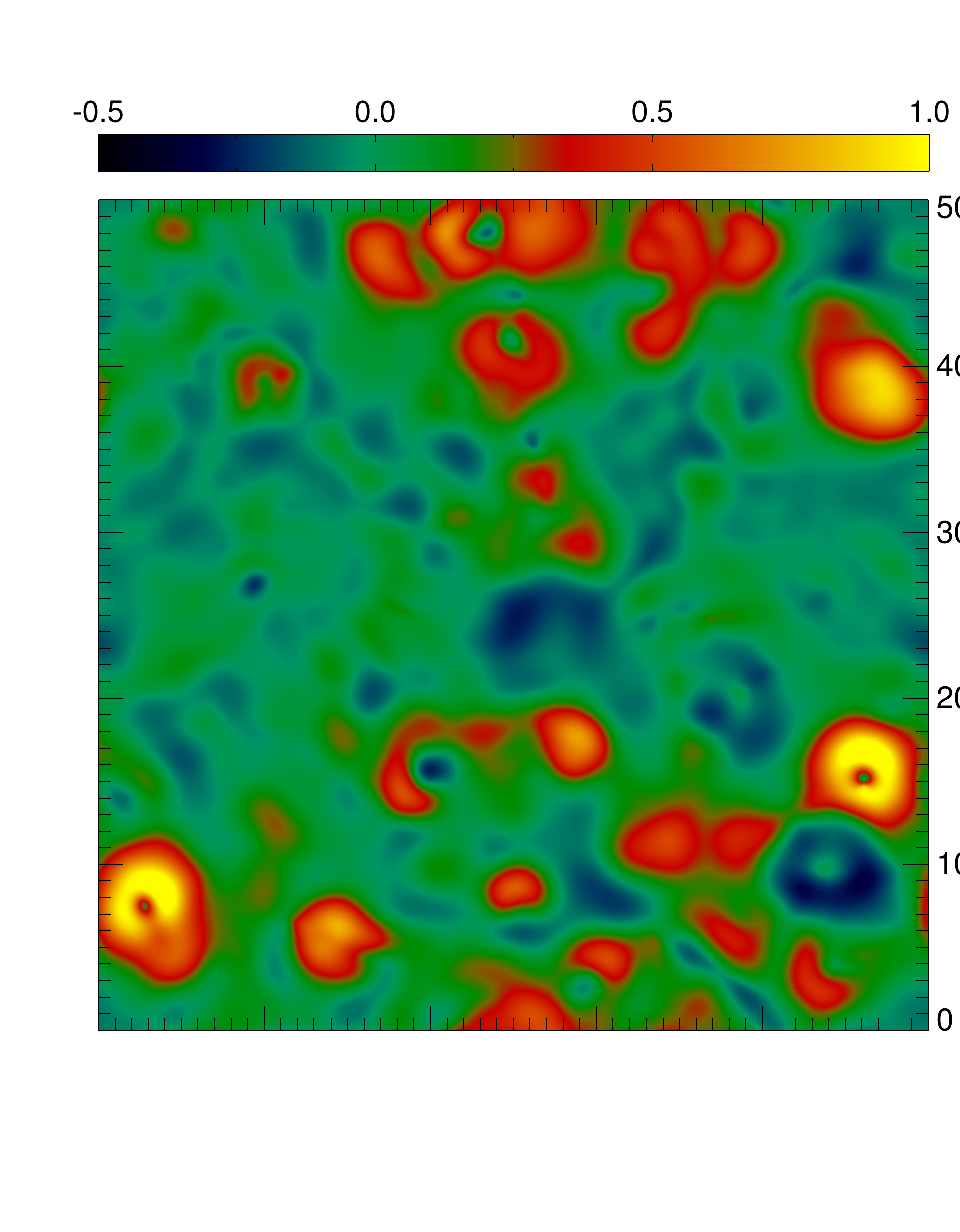}
\vspace{-2.9cm}
\\
\includegraphics[width=7.8cm]{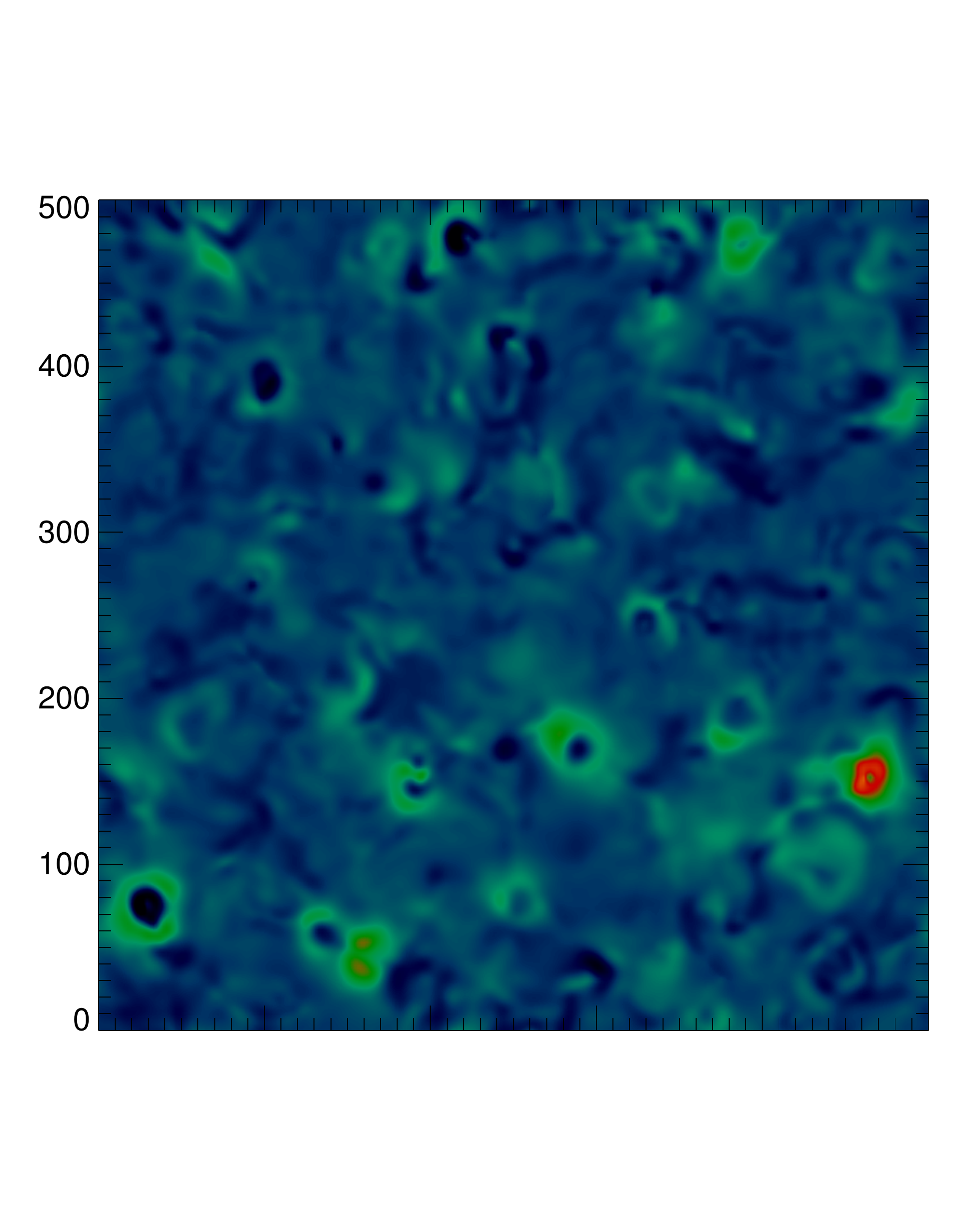}
\put(-225,140){\rotatebox[]{90}{{$Y$ [{\Muns}]}}}
\hspace{-.7cm}
\includegraphics[width=7.8cm]{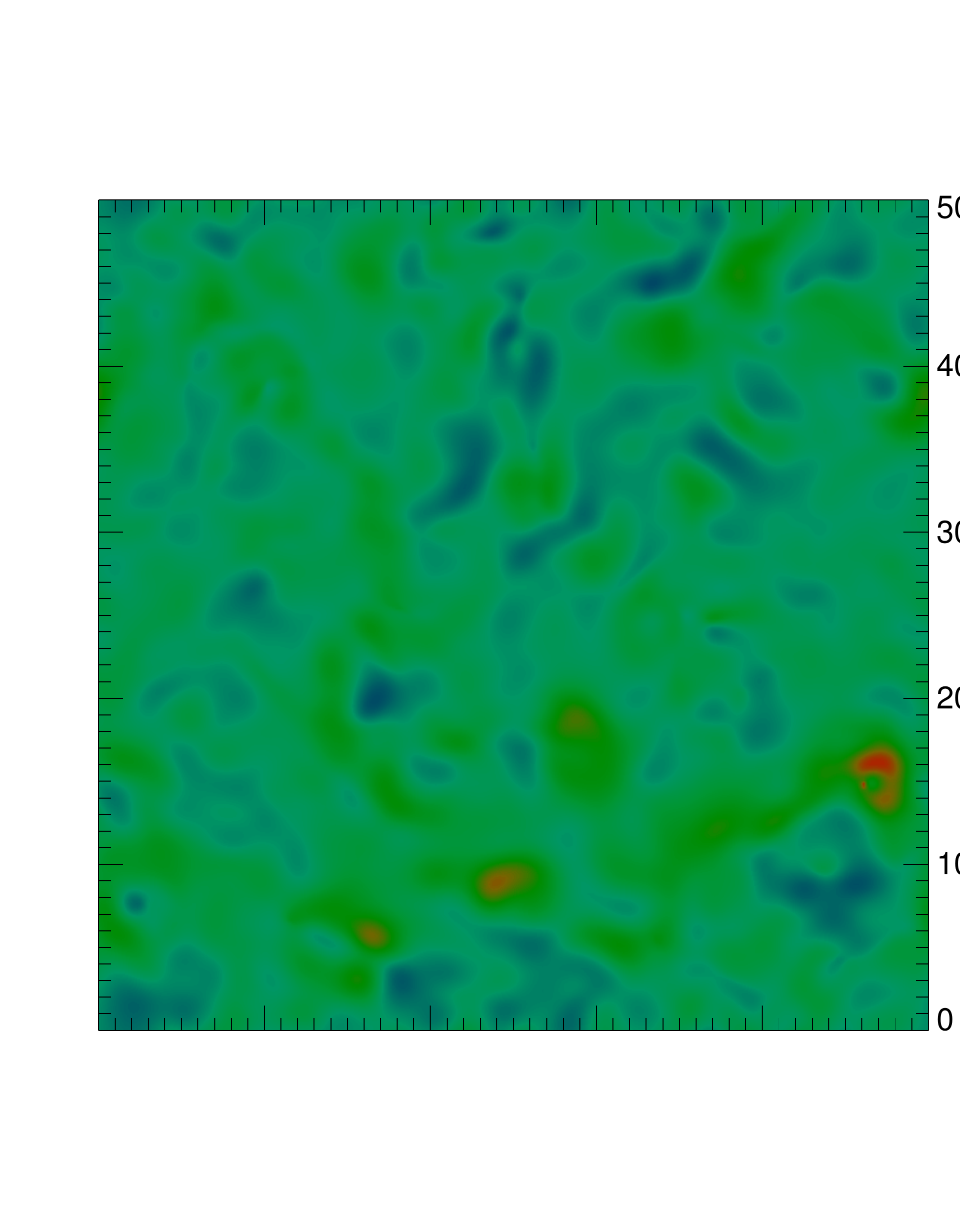}
\vspace{-2.9cm}
\\
\includegraphics[width=7.8cm]{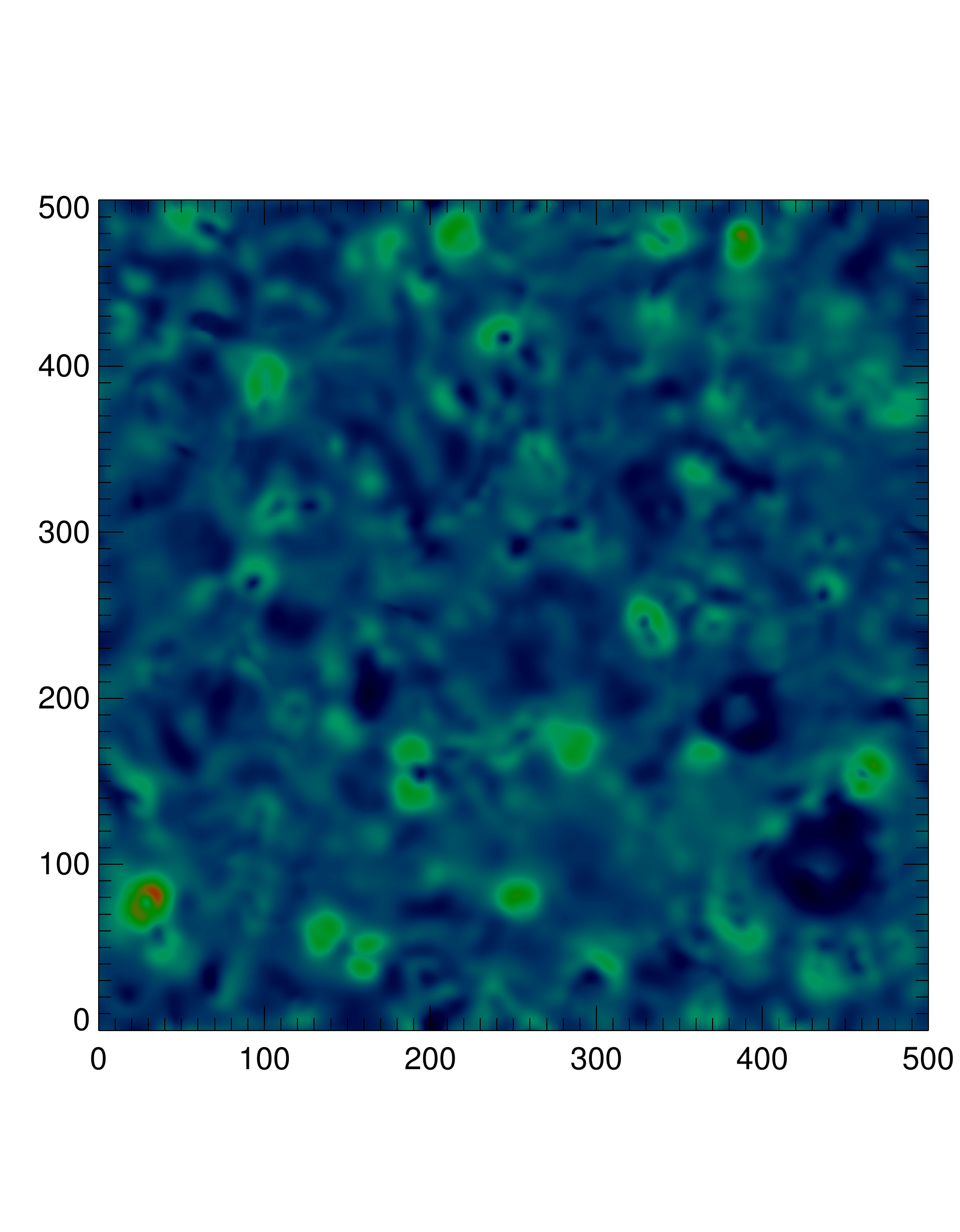}
\put(-225,140){\rotatebox[]{90}{{$Y$ [{\Muns}]}}}
\put(-130,25){{$X$ [{\Muns}]}}
\hspace{-.7cm}
\includegraphics[width=7.8cm]{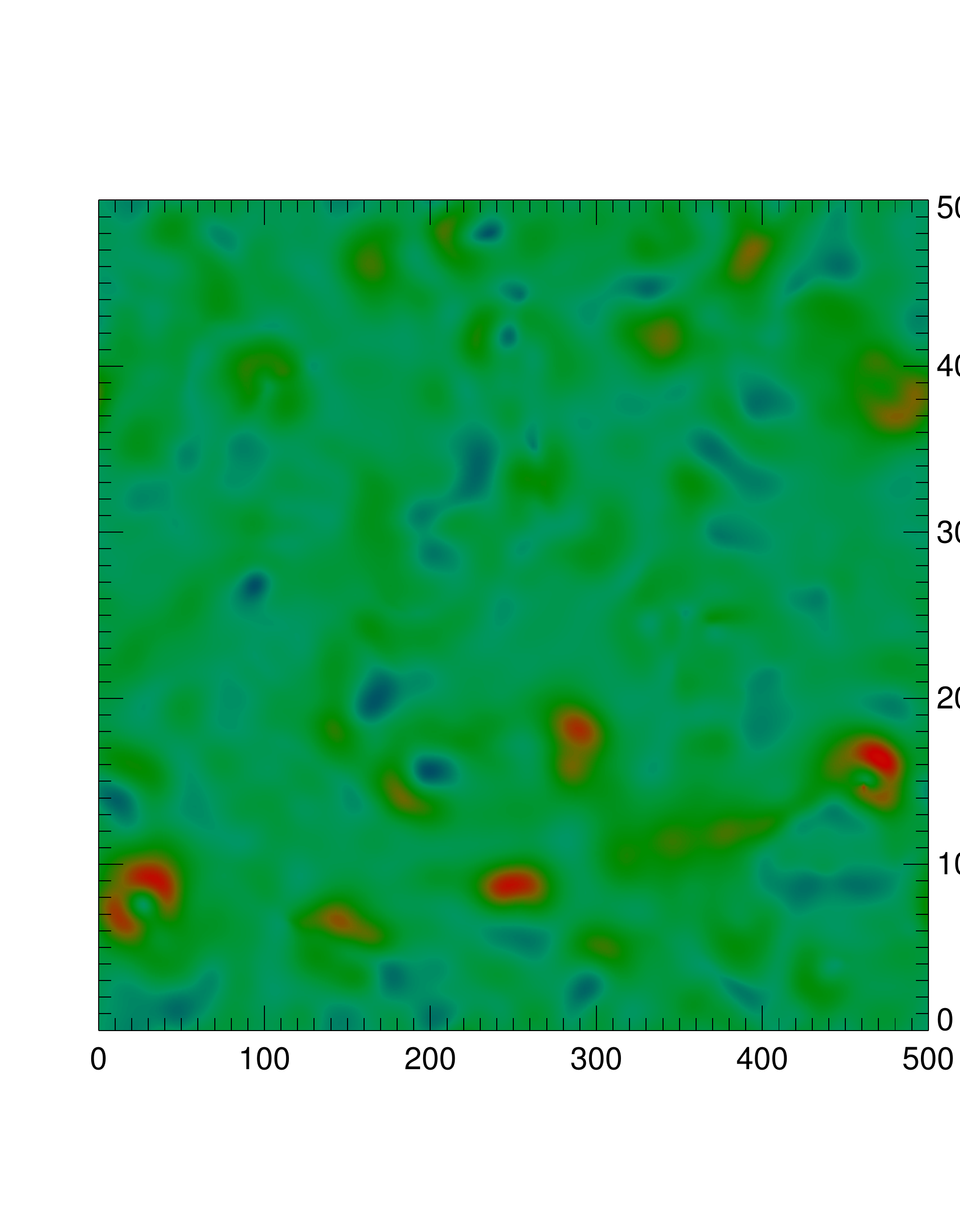}
\put(-130,25){{$X$ [{\Muns}]}}
\vspace{-1.cm}
\end{tabular}
\caption{\label{fig:corrtheta} Difference between the true and the estimated speeds  (velocity magnitude). On scales of $r_{\rm S}=$5 {\Muns} (left panels) and 10 {\Muns} (right panels) with different colour ranges. Upper panels: LIN, middle panels:  LOG-2LPT, lower panels: 2LPT. The colour-code is in units of $10^{-2}$ s$^{-1}$km.}
\end{figure*}

\begin{figure}
\begin{tabular}{cc}
\includegraphics[width=8.5cm]{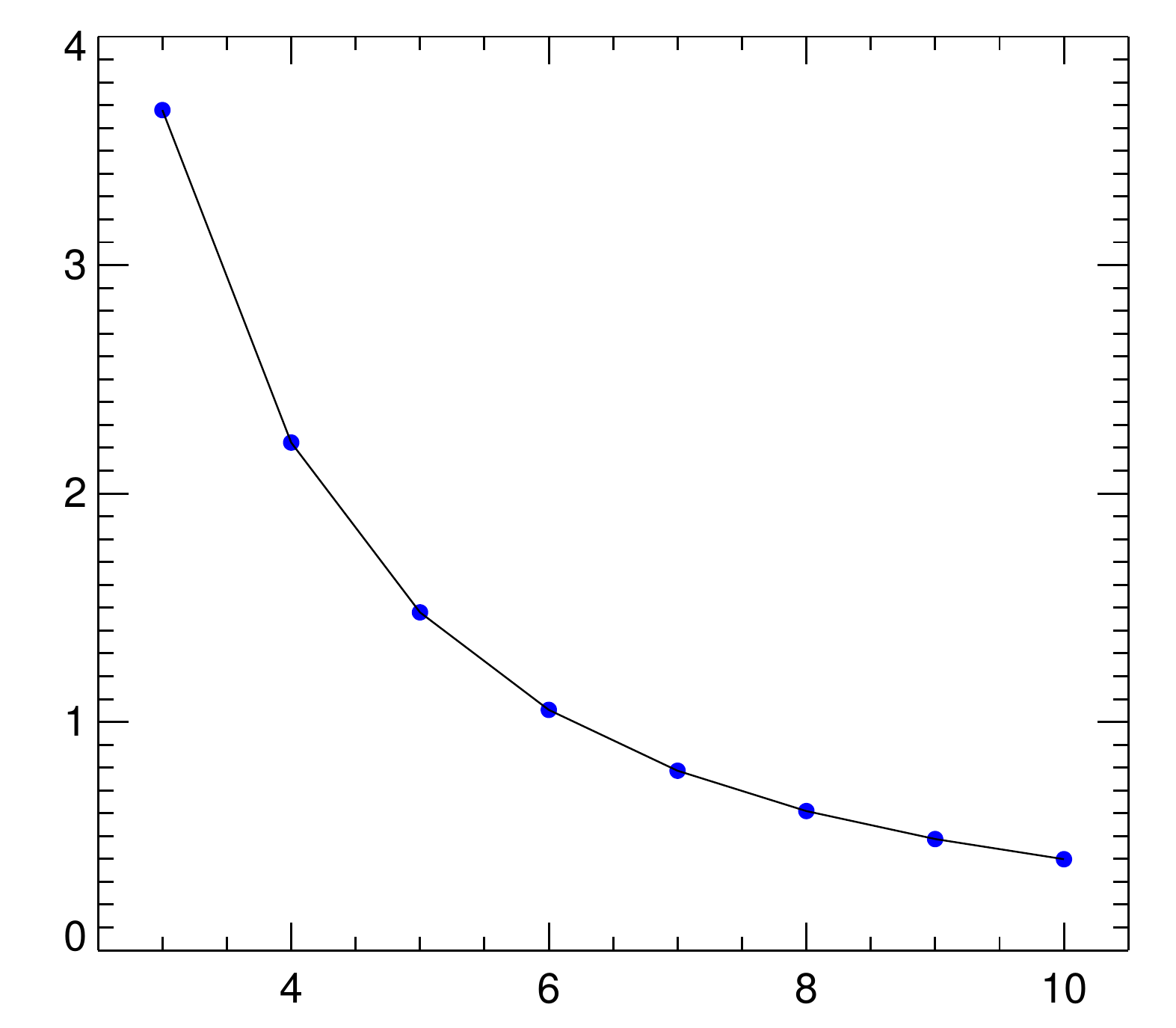}
\put(-240,105){\rotatebox[]{90}{{$\sigma(\nabla\times\mbi v|_x)/\sigma(\nabla\cdot\mbi v)$ [\%]}}}
\put(-140,-5){{$r_{\rm S}$ [{\Muns}]}}
\end{tabular}
\caption{\label{fig:sdev} Standard deviation of the $x$-component of the curl $\sigma(\nabla\times\mbi v|_x)$ divided by the standard deviation of the velocity divergence $\sigma(\nabla\cdot\mbi v)$ in \% as a function of the scale $r_{\rm S}$. The mean of both the curl and the divergence of the peculiar field are very close to zero. 
}
\end{figure}

The first ingredient in our algorithm is the linear component of the density
field. 	We stress that this field are not  the ``initial conditions''
of the universe, since structures have moved from its Lagrangian position to the
Eulerian ones at $z=0$.  Contrarily to the linear term, the probability distribution function (PDF) of the nonlinear component is highly skewed.

Fig.~\ref{fig:stats2} compares the PDF of the velocity divergence $\theta$ as
given by different estimators, with the one directly measured from the
simulation (blue dashed line). The left panel shows the fields smoothed with a
5 {\Muns} whereas the right panels do so with a larger smoothing of 10
{\Muns}.  In both cases, the predictions of linear perturbation theory
(i.e. $\theta=\delta$) displays the worst performance of all -- it overestimates
systematically the number of volume elements with large value of $\theta$ and
underestimates the ones with low values of $\theta$ (green curve). This is a
consequence of linear theory breaking down even in mildly under- or overdense
regions.

Using linear theory together with the linear component of the nonlinear density
field given by the logarithm of the field ($\ln(1+\delta_{\rm M})-\mu$, see
cyan curve in Fig.~\ref{fig:stats2}) is still a poor description of the PDF,
suggesting the need of higher order corrections. {\color{black} We note that this approximation corresponds to linear Lagrangian perturbation theory (Zeldovich approximation).}

{\color{black}  While standard linear (Eulerian) theory overestimates the values of the divergence of the peculiar velocity field, linear Lagrangian perturbation theory underestimates them. In the former case we are assuming that the scaled divergence of the peculiar velocity field is given by the density field at (final) Eulerian coordinates ($\theta=\delta(\mbi x)$), i.~e. by the full nonlinear density field; whereas in the latter case we assume that it is given by the density field at (initial) Lagrangian coordinates transformed to Eulerian coordinates ($\theta=D_1\delta^{(1)}(\mbi x)$), i.~e. by the first linear (Gaussian) term in Lagrangian perturbation theory. We note that in general the underestimation of the Zeldovich approximation is less severe (cyan curve) than the overestimation of linear theory (green curve) in high density regions (see Fig.~\ref{fig:stats2}). Meaning that the Zeldovich approximation performs better than linear theory being more conservative, as it has been repeatedly shown in the literature (see e.g. Nusser and Branchini 2000).
 Velocity reconstruction approaches like PIZA or MAK are based on the Zeldovich approximation \citep[see][respectively]{1997MNRAS.285..793C,LMCTBS08}. We thus expect that the approach presented here is more accurate than these methods.}

The first second order estimation we consider is that given by
Eq.~(\ref{eq:2lpttheta3}) which is closely related to the one proposed by
Gramann and uses the nonlinear field as a proxy for the linear density field
(yellow curve). In the regime where this approximation is valid ($\delta \sim
0$) this approach performs remarkably well. However, there is a clear and rapid
degradation for volume elements with larger deviations of homogeneity. For instance,
this solution yields to values even of $\theta=-140$ at scales of 5
{\Muns}!  This behaviour is due to a complete misestimation of the
nonlinear term $\delta^{(2)}$ and therefore of the nonlinear corrections to the
velocity field. 

Finally, black and red lines indicate the results of the method proposed in
this paper: Lagrangian perturbation theory based upon an estimate of the
linear component of the density field using the lognormal model (LOG-2LPT:
black) or based on the iterative Lagragian linearisation (2LPT: red). In both panels,
the predicted PDF very closely follows that measured in the Millennium
simulation, even on the extreme tails (especially with LOG-2LPT). The only
appreciable difference with the lognormal model is a slight overestimation for
low values of $\theta$ (static regions), we note however, that this could be
potentially improved by higher order expressions.  Indeed, 3rd order PT appears
to perform better than 2LPT for underdense regions (see appendix).  In spite of
this, on both 5 and 10 {\Muns} \footnote{We have also checked that this is
true on 3,4,6,7 and 8 {\Muns}} our method is clearly superior to any of the
other methods we investigated here, as far as the PDF is concerned and for any
value of $\theta$.  The iterative 2LPT solution yields a moderate
overestimation for high values of $\theta$ due to the laminar flow
approximation used in Lagrangian perturbation theory which does not fully
capture nonlinear structure formation. Nevertheless, the PDF of $\theta$ using
this solution  is clearly superior to the linear approximation.

We now continue with a more detailed testing of our method. In
Fig.~\ref{fig:thetadiff} we plot the predicted velocity divergence,
$\theta^{\rm rec}_{\rm LOG-2LPT}$ and $\theta^{\rm rec}_{\rm 2LPT}$ (based on
$\delta^{\rm L}_{\rm LOG}$  and $\delta^{\rm L}_{\rm LPT}$, respectively), for
each of the $256^3$ cells in our mesh, as a function of the value measured in
the simulation. As in the previous plot, we display results on two different
scales; 5 {\Muns} on the left panels and 10 {\Muns} on the right panels.
For comparison we also provide the results using linear perturbation theory
$\theta^{\rm rec}_{\rm LIN}=\delta$. 

The values of $\theta$ are remarkably well predicted by Lagrangian perturbation
theory.  In fact, measurements lie around the 1:1 line in the LOG-2LPT case,
implying that there are no appreciable biases in our estimation over all the
range probed by the Millennium simulation (with the exception of the low values
of $\theta$, for an improvement on this see appendix). 
In contrast, the linear theory prediction presents overestimations of up to a
factor of 3 for the 5 {\Muns} smoothing and of 2 for 10 {\Muns}. 

The iterative solution (2LPT) produces smaller dispersions but also a
slight overestimation of $\theta$ for high values, as we already mentioned
before.

The distribution of differences in our method is well approximated by a
Gaussian function, whereas in linear theory there are significant extended
tails, we will return to this in more detail in \S3.2.  Overall, this plot
suggest that our method not only performs adequately on a statistical basis,
but also on predicting the actual average value of $\theta$ in a given volume
element.

Although not displayed by the figure, our method also performs better than the
other methods shown in Fig.~\ref{fig:stats2}.  In particular, the classic
application of 2LPT recovers $\theta$ quite well for the range $-0.5<\theta<2$,
as shown in \citet{vdG93}. However, outside this range it displays an erratic
behaviour as it could have been anticipated from Fig.~\ref{fig:stats2}.

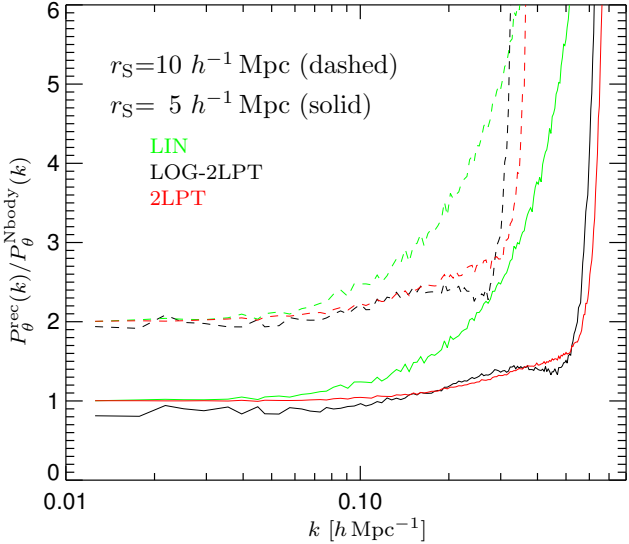
\begin{figure}
\begin{tabular}{cc}
\includegraphics[width=8.5cm]{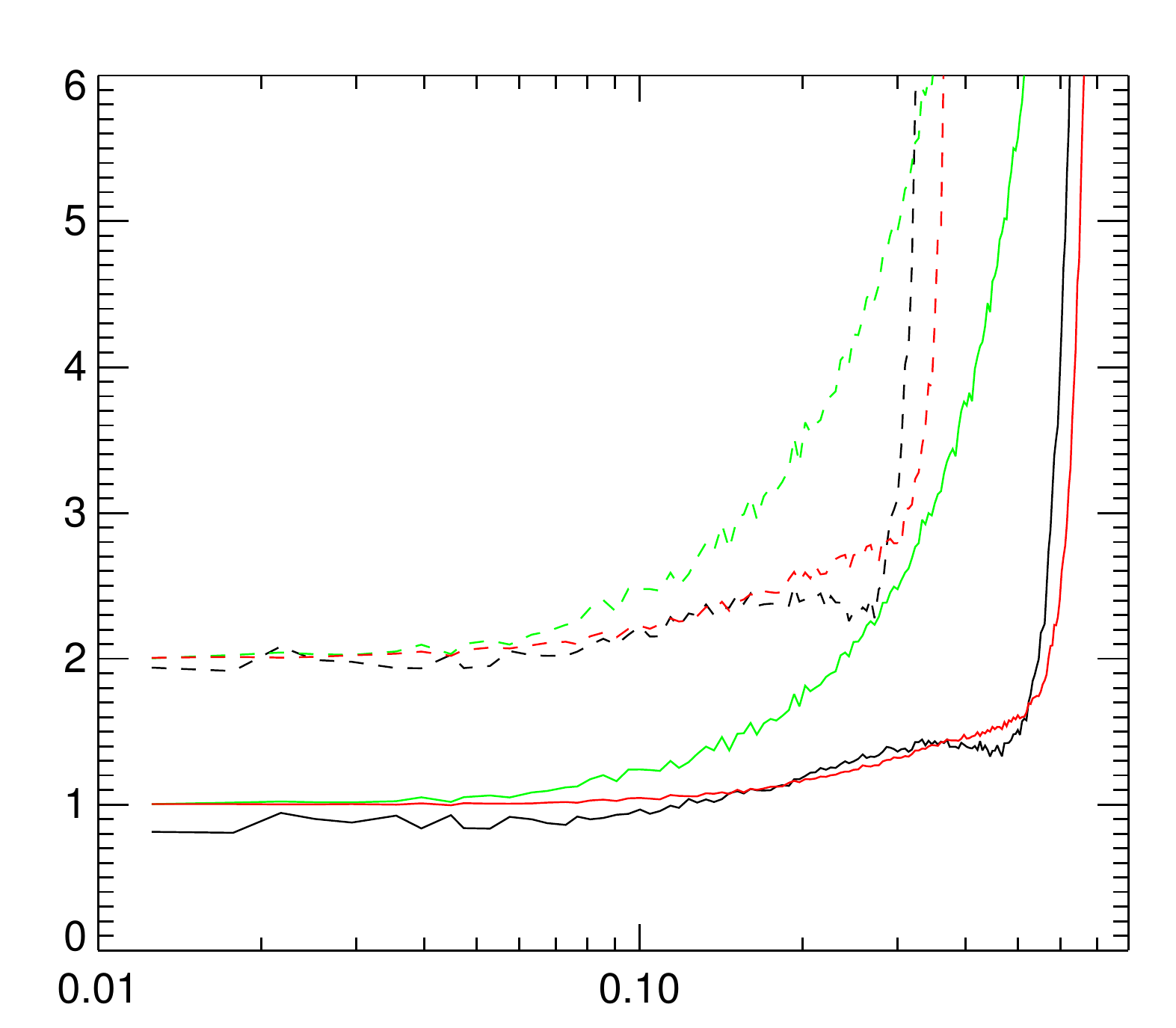}
\put(-205,170){\large $r_{\rm S}$=10 {\Muns} (dashed)}
\put(-205,155){\large $r_{\rm S}$=\,\,\,5 {\Muns} (solid)}
\put(-190,140){\color{green}LIN}
\put(-190,130){\color{black}LOG-2LPT}
\put(-190,120){\color{red}2LPT}
\put(-130,-5){{$k$ [{\kuns}]}}
\put(-245,100){\rotatebox[]{90}{{$P^{\rm rec}_\theta(k)/P^{\rm Nbody}_\theta(k)$}}}
\end{tabular}
\caption{\label{fig:pow} Ratio of the true $P^{\rm Nbody}_\theta(k)$ and reconstructed $P^{\rm rec}_\theta(k)$ power-spectra of the scaled velocity divergence $\theta$ for green: linear theory (LIN), black: 2LPT estimate from the logarithm of the density field (LOG-2LPT) (Eq.~\ref{eq:2lpttheta}), red:  2LPT estimate from the iterative solution (2LPT) (see \S\ref{sec:method}). {\color{black} Note that in the case of 10 {\Muns} smoothing we have multiplied the lines by a factor of 2 for clarity.}
}
\end{figure}

A complementary visual assessment of our method is provided in
Fig.~\ref{fig:corrtheta}. In these images we display the relative difference
$\theta^{\rm rec}-\theta^{\rm Nbody}$ ($\theta^{\rm rec}_{\rm LIN}-\theta^{\rm
Nbody}$ in the upper panels, $\theta^{\rm rec}_{\rm 2LPT}-\theta^{\rm Nbody}$
in the central panels and $\theta^{\rm rec}_{\rm LOG-2LPT}-\theta^{\rm Nbody}$
in the lower ones) projected on a slice 2 {\Muns} thick.  As previously, we
explore 5 and 10 {\Muns} and also display the linear theory predictions for
comparison. We see that this difference field is not uniformly distributed
across the simulation but there are well defined regions in which the
prediction is very accurate and others where the prediction is somewhat
worse.  Not surprisingly the latter coincide with high density regions.
Nevertheless, and consistently with previous plots, we see that areas where
linear theory fails dramatically are much better handled in our approach. 

As a final crucial check, we compute the power-spectra of the scaled velocity
divergence according to the $N$-body simulation and our reconstruction with
both the LPT and the lognormal linearisation approaches. This is shown in
Fig.~\ref{fig:pow}. Linear theory, as expected,  increasingly overestimates the
power towards small scales, whereas the 2LPT solutions peform remarkably well
in a wide range going down to scales of {\color{black} $k\simeq0.3\,h$/Mpc and $k\simeq0.5\,h$/Mpc for $r_{\rm S}$=10 {\Muns} and $r_{\rm S}$=5 {\Muns}, respectively}. We can also see that
there is a systematic deviation originated by the lognormal transformation.
The LPT estimate of the linear component corrects these and the results are
extremely close to the actual power-spectrum over most of the $k$-range shown.
\subsection{The full 3D peculiar velocity field} \label{sec:vres}

\begin{figure*}
\begin{tabular}{cc}
\includegraphics[width=8.cm]{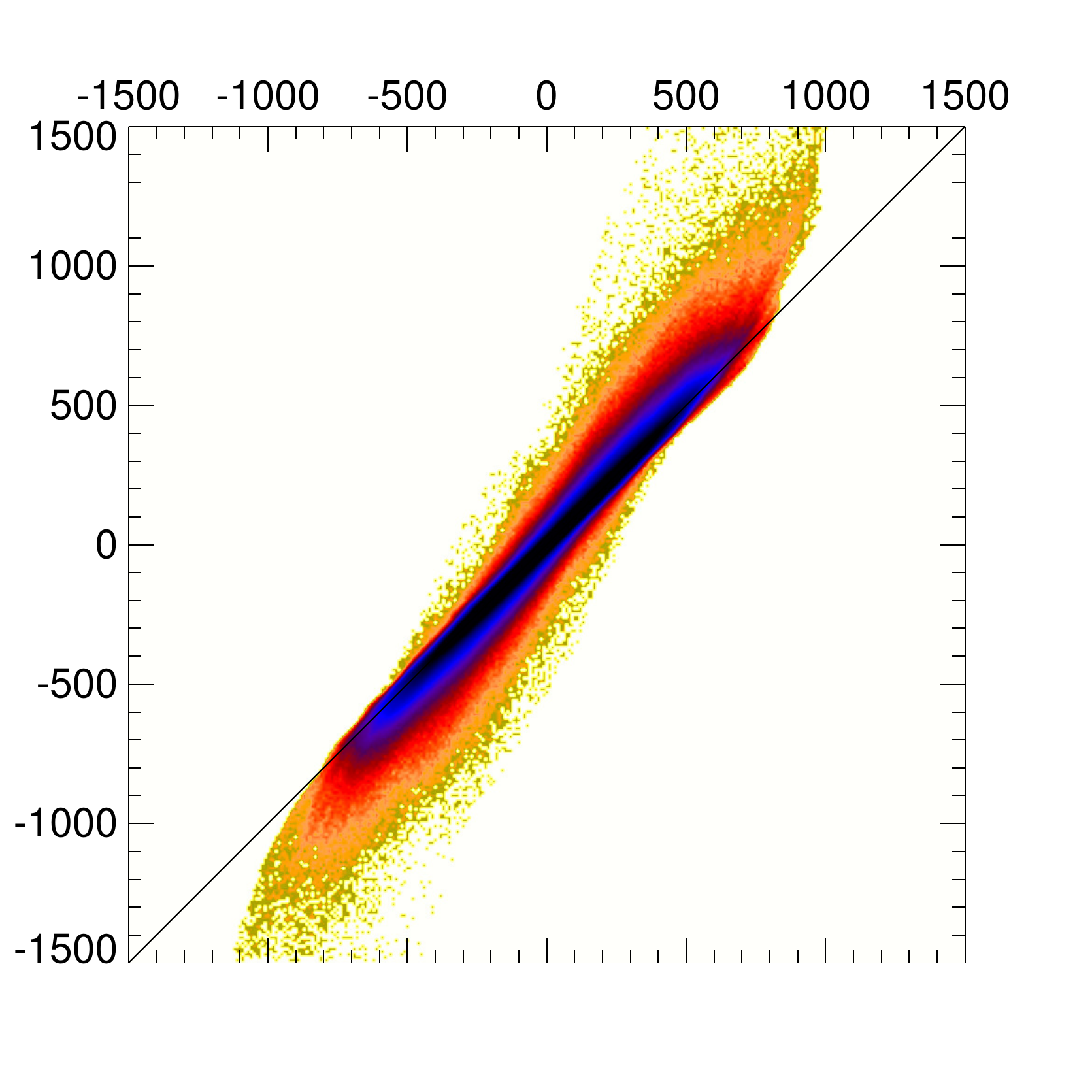}
\put(-100,70){{\large $r_{\rm S}$=5 {\Muns}}}
\put(-100,50){{\large LIN}}
\put(-235,110){\rotatebox[]{90}{{$v_x^{\rm rec}$ [s$^{-1}$km]}}}
\hspace{-1.5cm}
\includegraphics[width=8.cm]{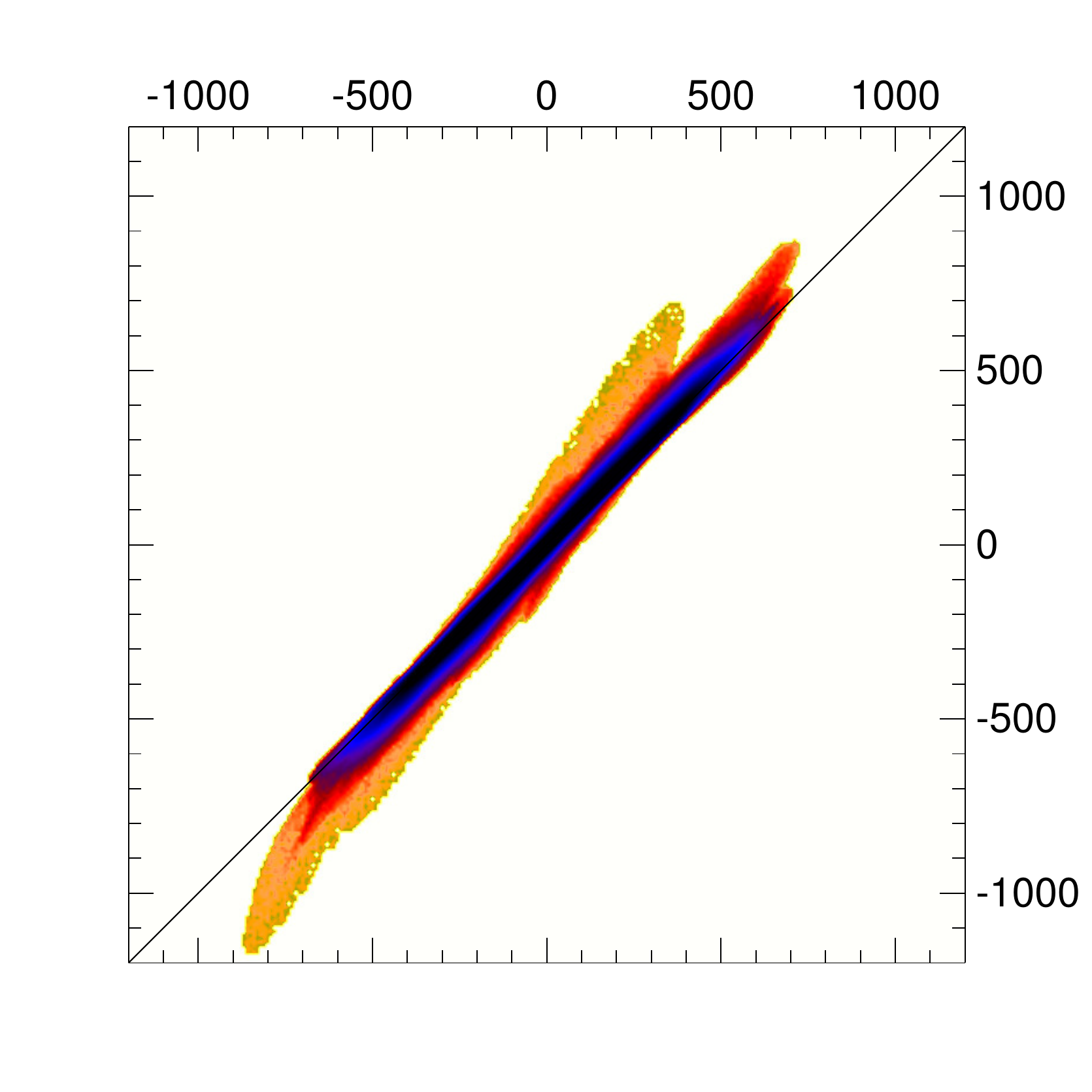}
\put(-100,70){{\large $r_{\rm S}$=10 {\Muns}}}
\put(-100,50){{\large LIN}}
\vspace{-1.5cm}
\\
\includegraphics[width=8.cm]{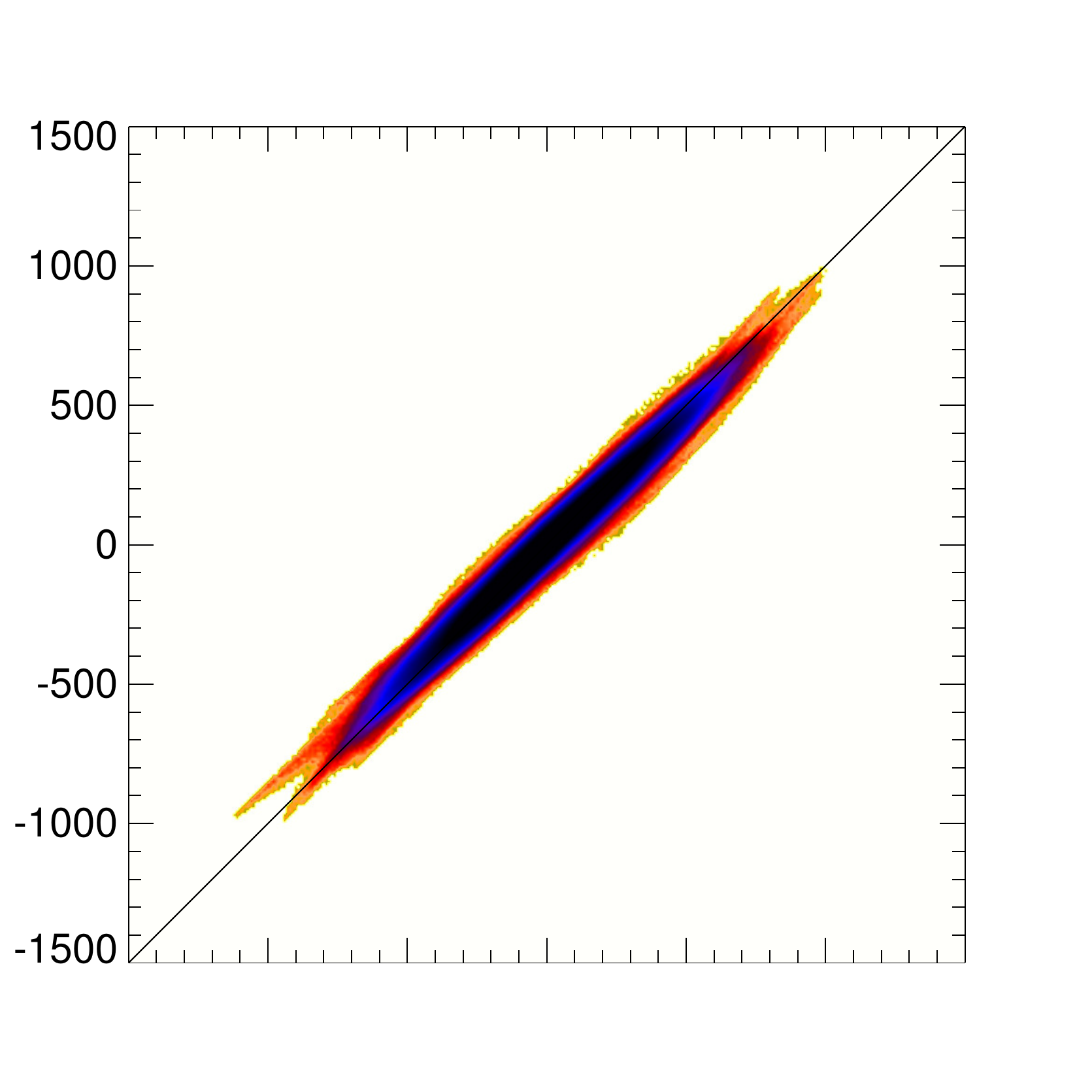}
\put(-235,110){\rotatebox[]{90}{{$v_x^{\rm rec}$ [s$^{-1}$km]}}}
\put(-100,70){{\large $r_{\rm S}$=5 {\Muns}}}
\put(-100,50){{\large LOG-2LPT}}
\hspace{-1.5cm}
\includegraphics[width=8.cm]{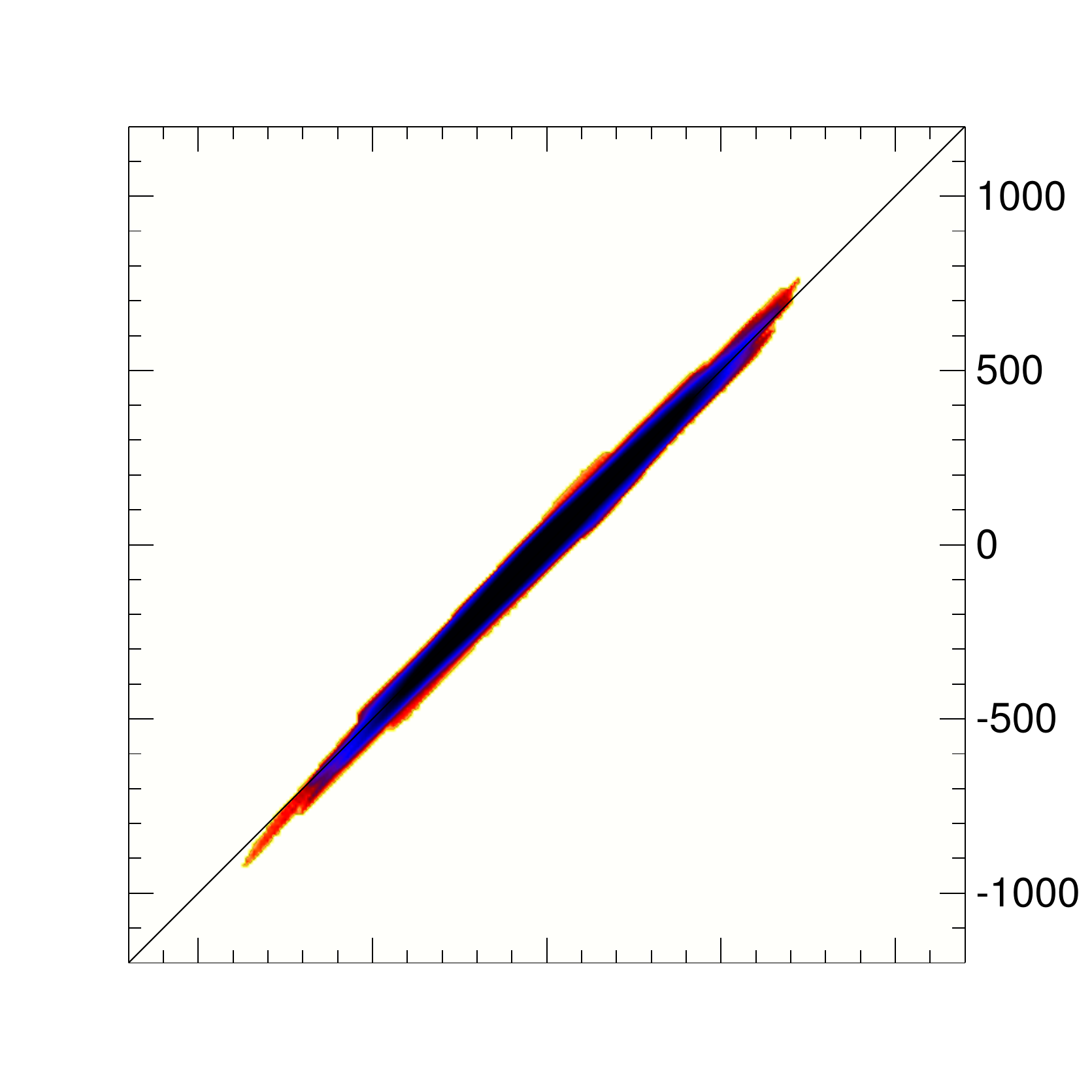}
\put(-100,70){{\large $r_{\rm S}$=10 {\Muns}}}
\put(-100,50){{\large LOG-2LPT}}
\vspace{-1.5cm}
\\
\includegraphics[width=8.cm]{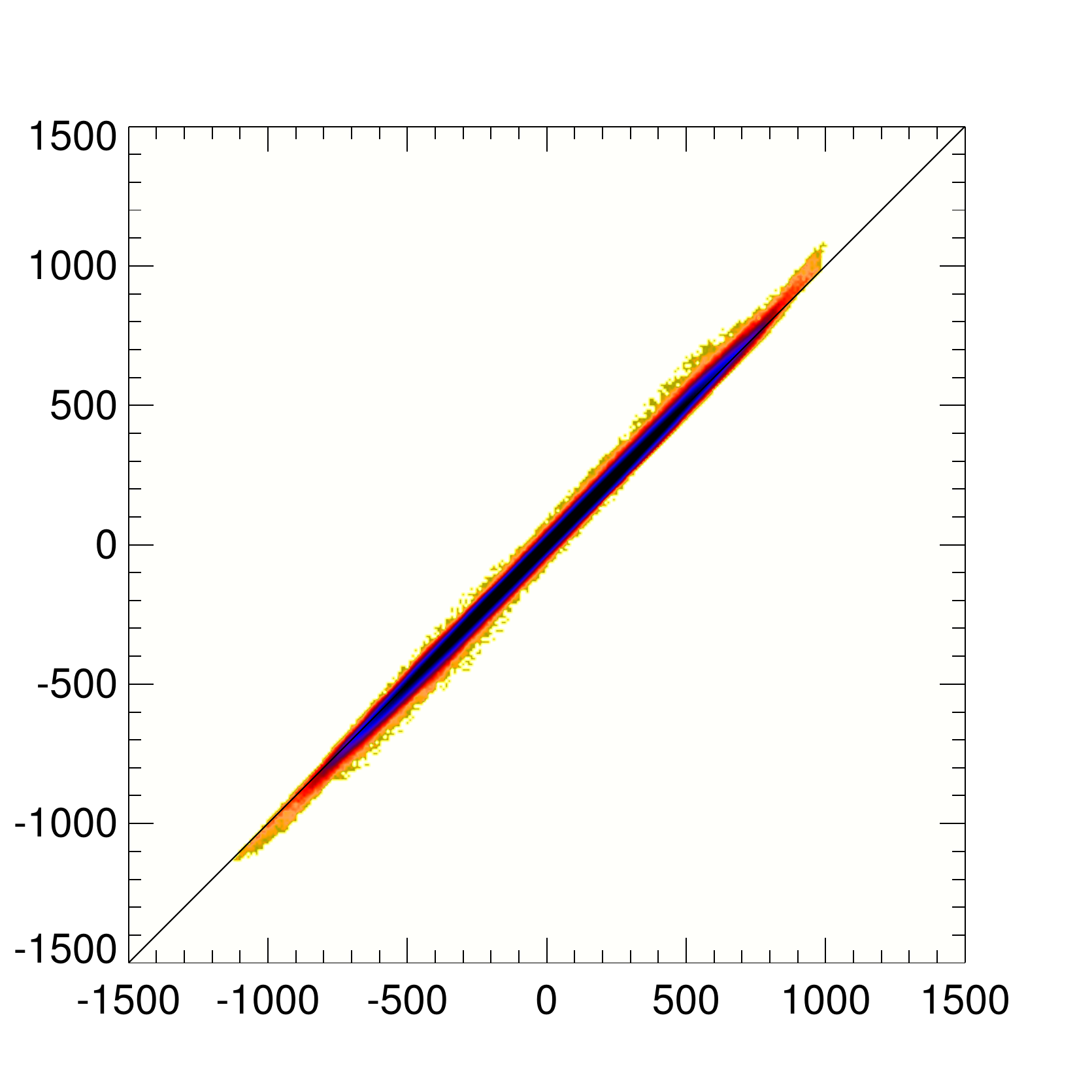}
\put(-235,110){\rotatebox[]{90}{{$v_x^{\rm rec}$ [s$^{-1}$km]}}}
\put(-100,70){{\large $r_{\rm S}$=5 {\Muns}}}
\put(-100,50){{\large 2LPT}}
\put(-125,0){{$v_x^{\rm Nbody}$ [s$^{-1}$km]}}
\hspace{-1.5cm}
\includegraphics[width=8.cm]{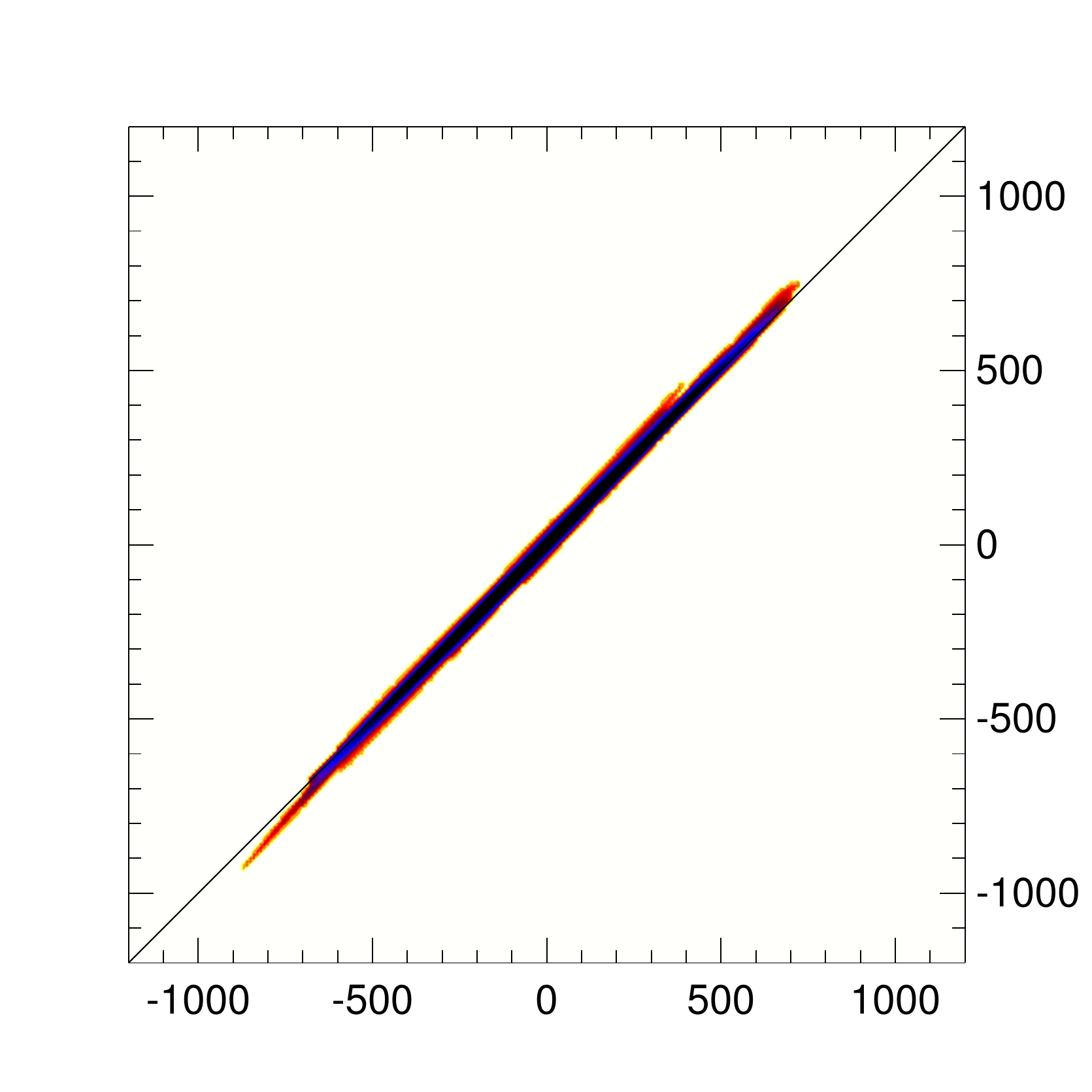}
\put(-100,70){{\large $r_{\rm S}$=10 {\Muns}}}
\put(-100,50){{\large 2LPT}}
\put(-125,0){{$v_x^{\rm Nbody}$ [s$^{-1}$km]}}
\vspace{0.cm}
\end{tabular}
\caption{\label{fig:veldiff} Cell-to-cell comparison between the true velocity $v_x^{\rm Nbody}$ and the reconstructed one $v^{\rm rec}_x$  for the $x$-component. Left panels on scales of 5 {\Muns} and right panels on scales of 10 {\Muns}. Upper panels: LIN, middle panels: LOG-2LPT, lower panels: 2LPT (see \S \ref{sec:method}). }
\end{figure*}

\begin{figure*}
\begin{tabular}{cc}
\includegraphics[width=8.5cm]{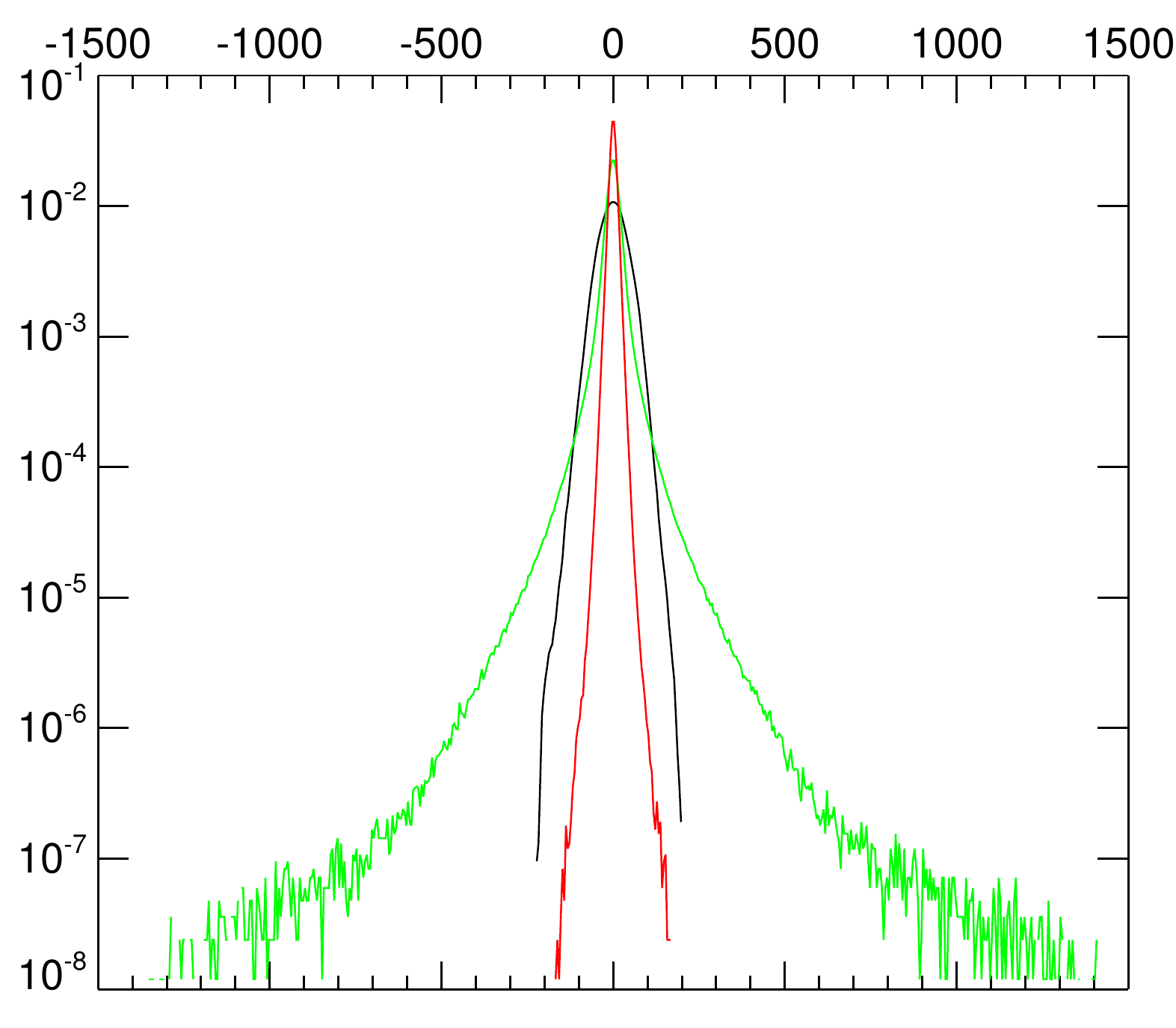}
\put(-85,165){{\large $r_{\rm S}$=5 {\Muns}}}
\put(-250,105){\rotatebox[]{90}{{PDF}}}
\put(-70,140){\color{green}LIN}
\put(-70,130){\color{black}LOG-2LPT}
\put(-70,120){\color{red}2LPT}
\includegraphics[width=8.5cm]{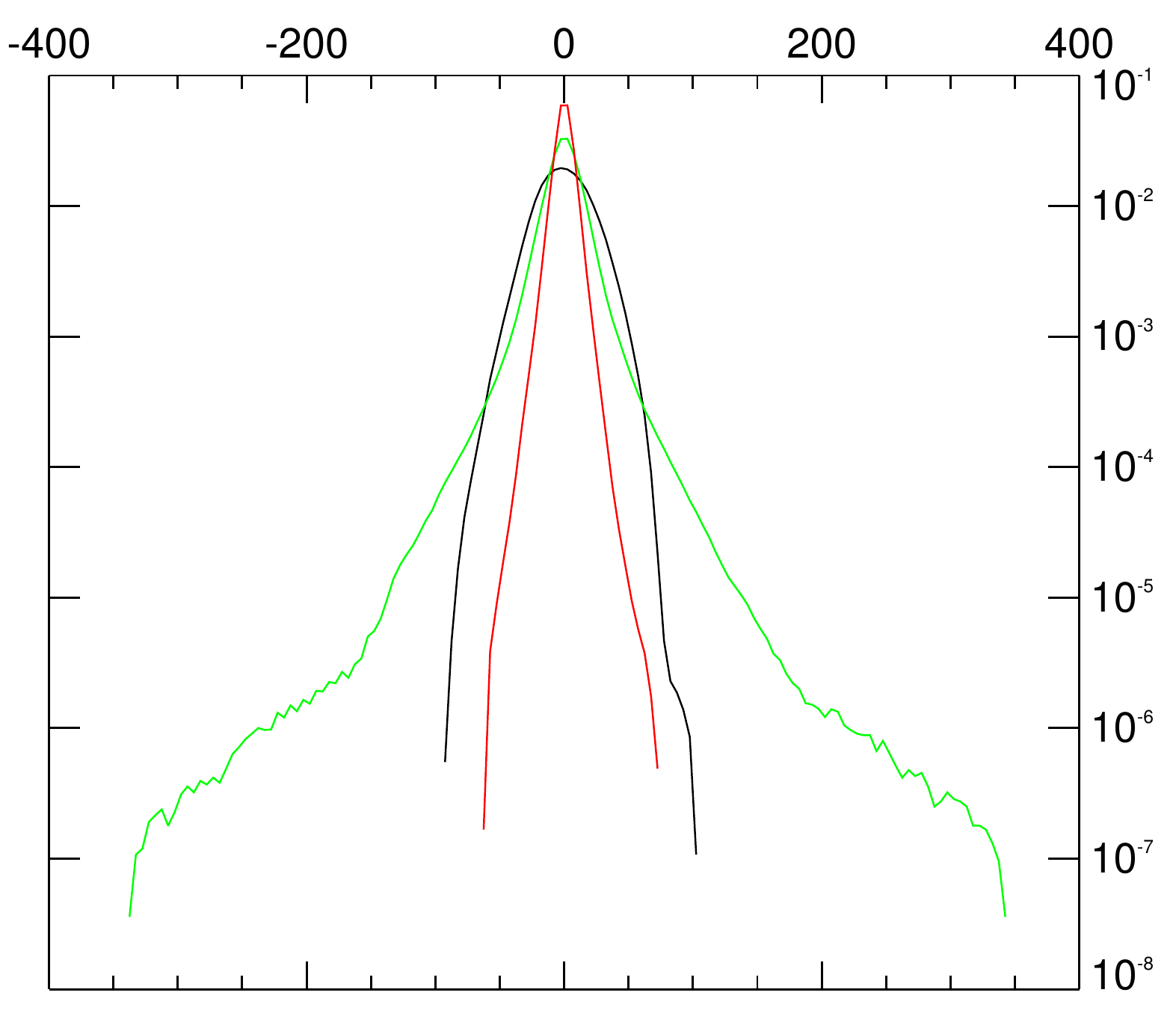}
\put(-95,165){{\large $r_{\rm S}$=10 {\Muns}}}
\\
\includegraphics[width=8.5cm]{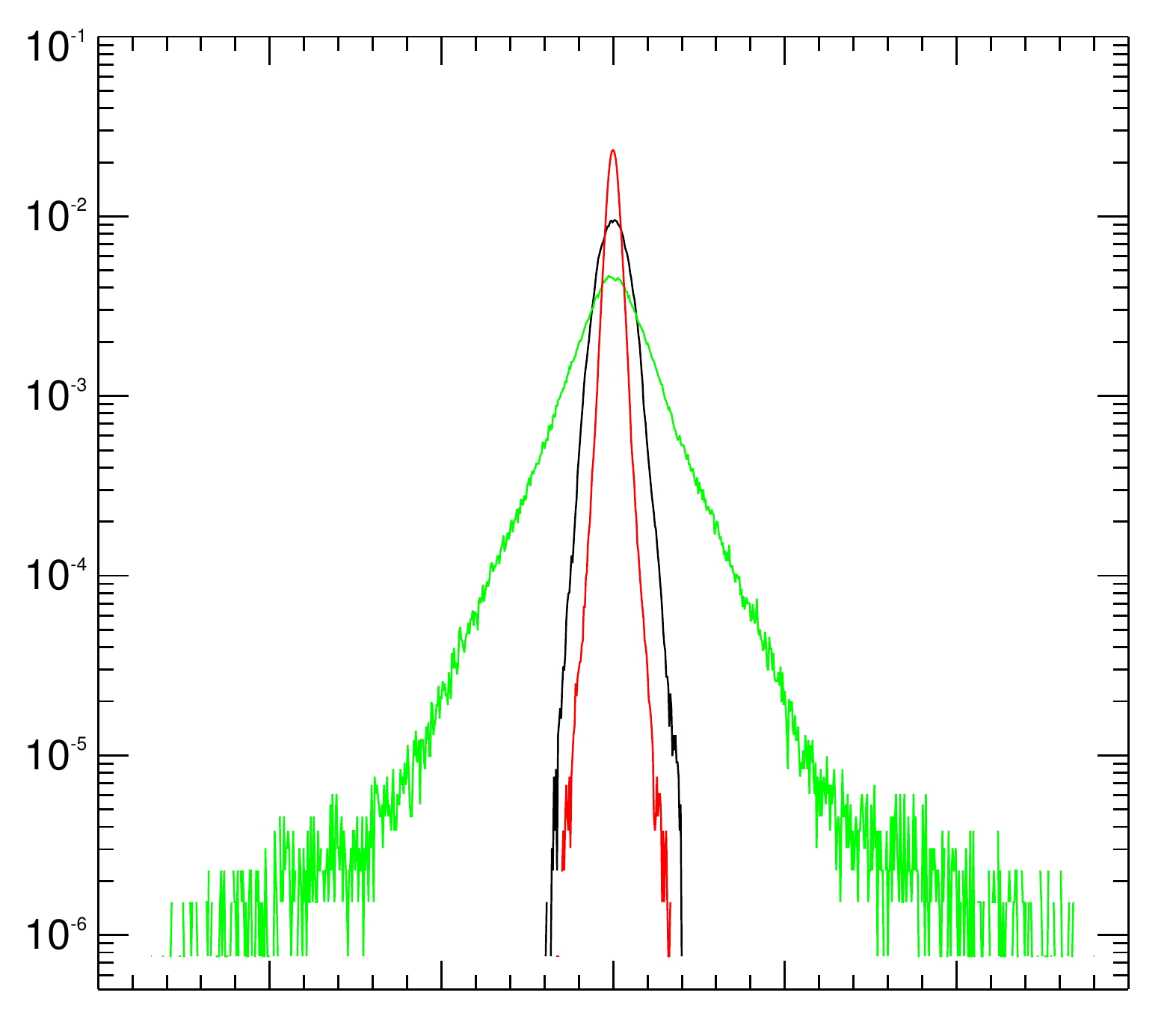}
\put(-250,105){\rotatebox[]{90}{{PDF}}}
\put(-180,165){{\large $\delta>2$}}
\put(-85,165){{\large $r_{\rm S}$=5 {\Muns}}}
\includegraphics[width=8.5cm]{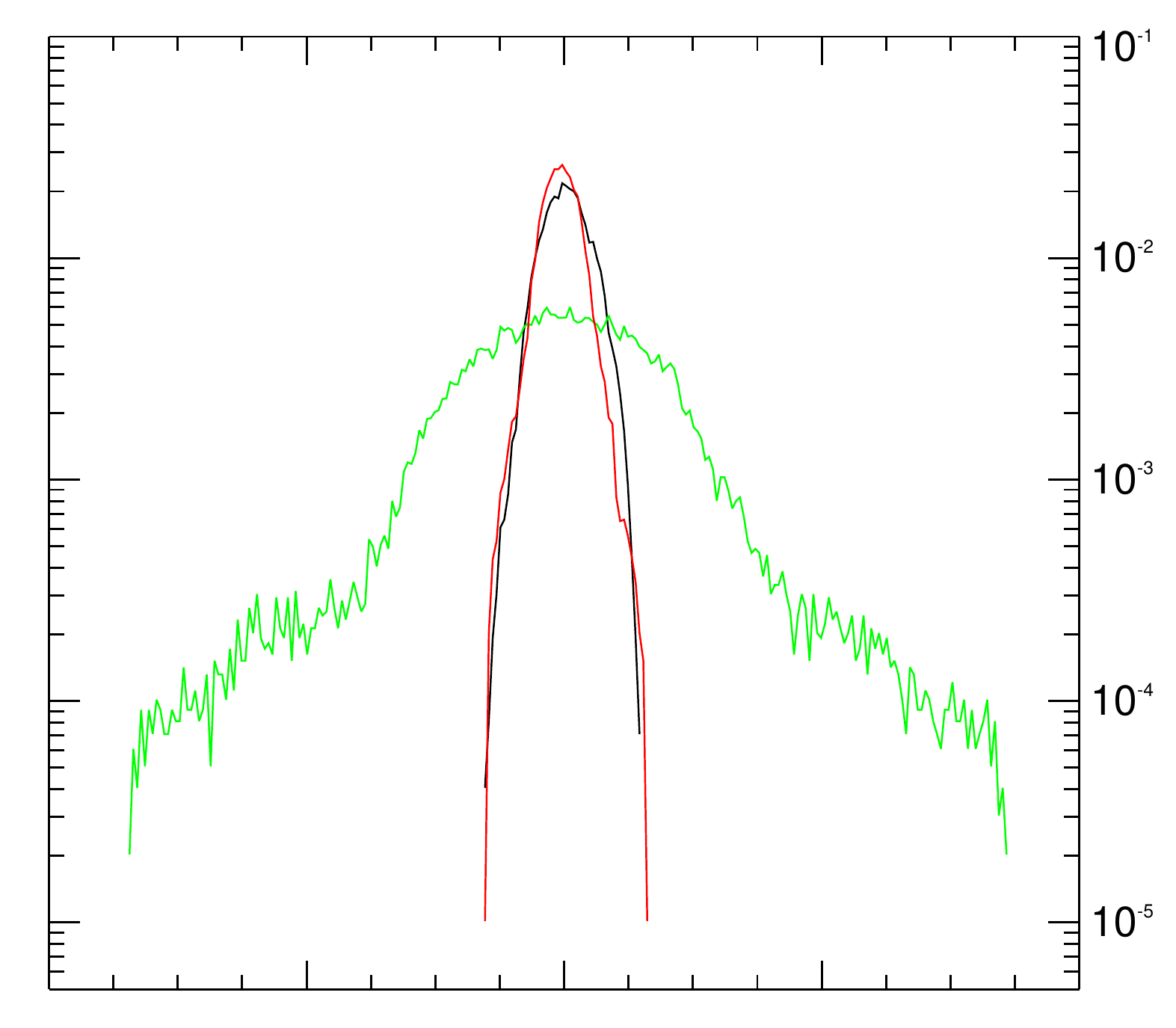}
\put(-180,165){{\large $\delta>2$}}
\put(-95,165){{\large $r_{\rm S}$=10 {\Muns}}}
\\
\includegraphics[width=8.5cm]{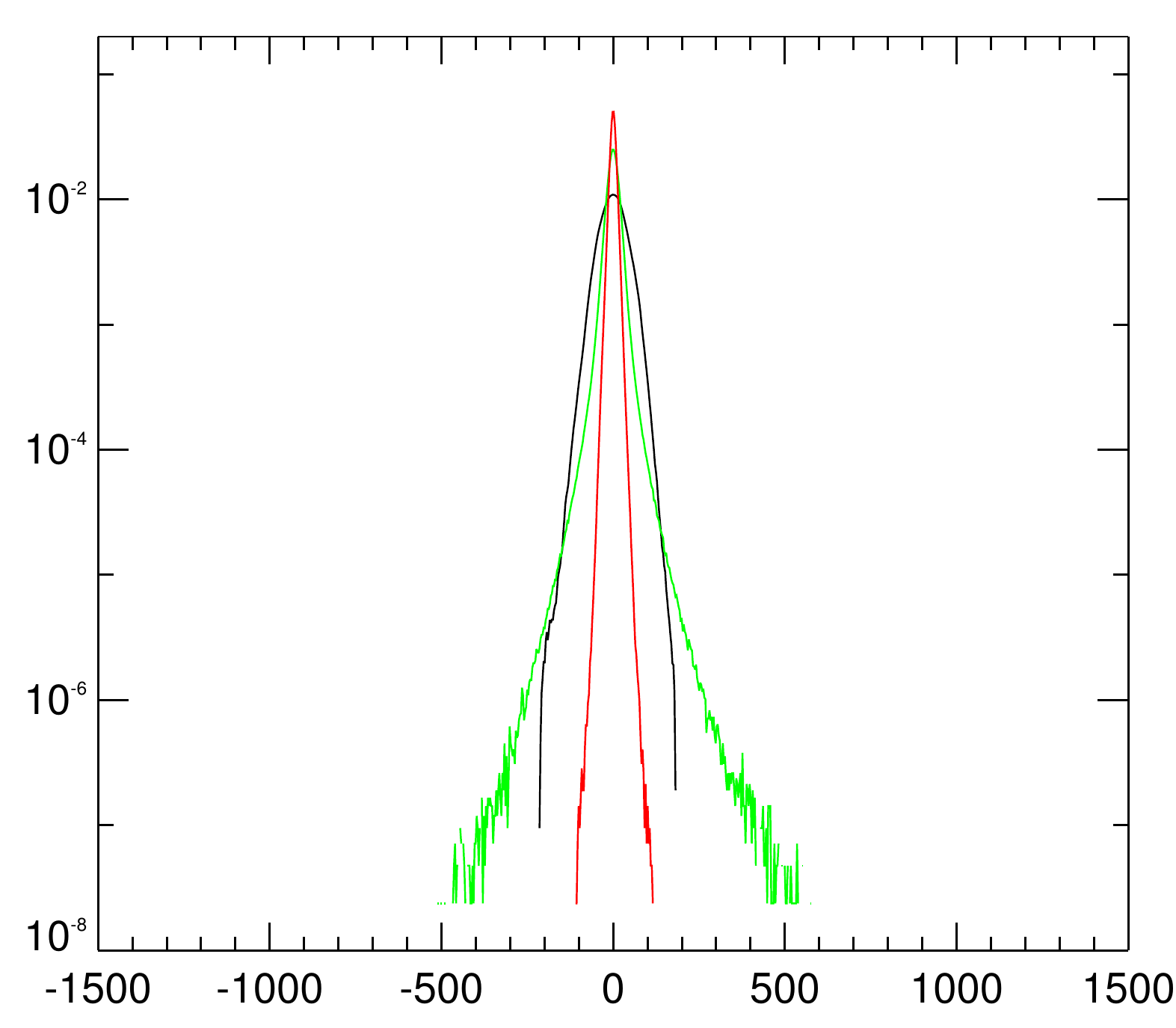}
\put(-250,105){\rotatebox[]{90}{{PDF}}}
\put(-123,-5){{$\epsilon_{v,x}$ [s$^{-1}$km]}}
\put(-180,165){{\large $\delta<0.5$}}
\put(-85,165){{\large $r_{\rm S}$=5 {\Muns}}}
\includegraphics[width=8.5cm]{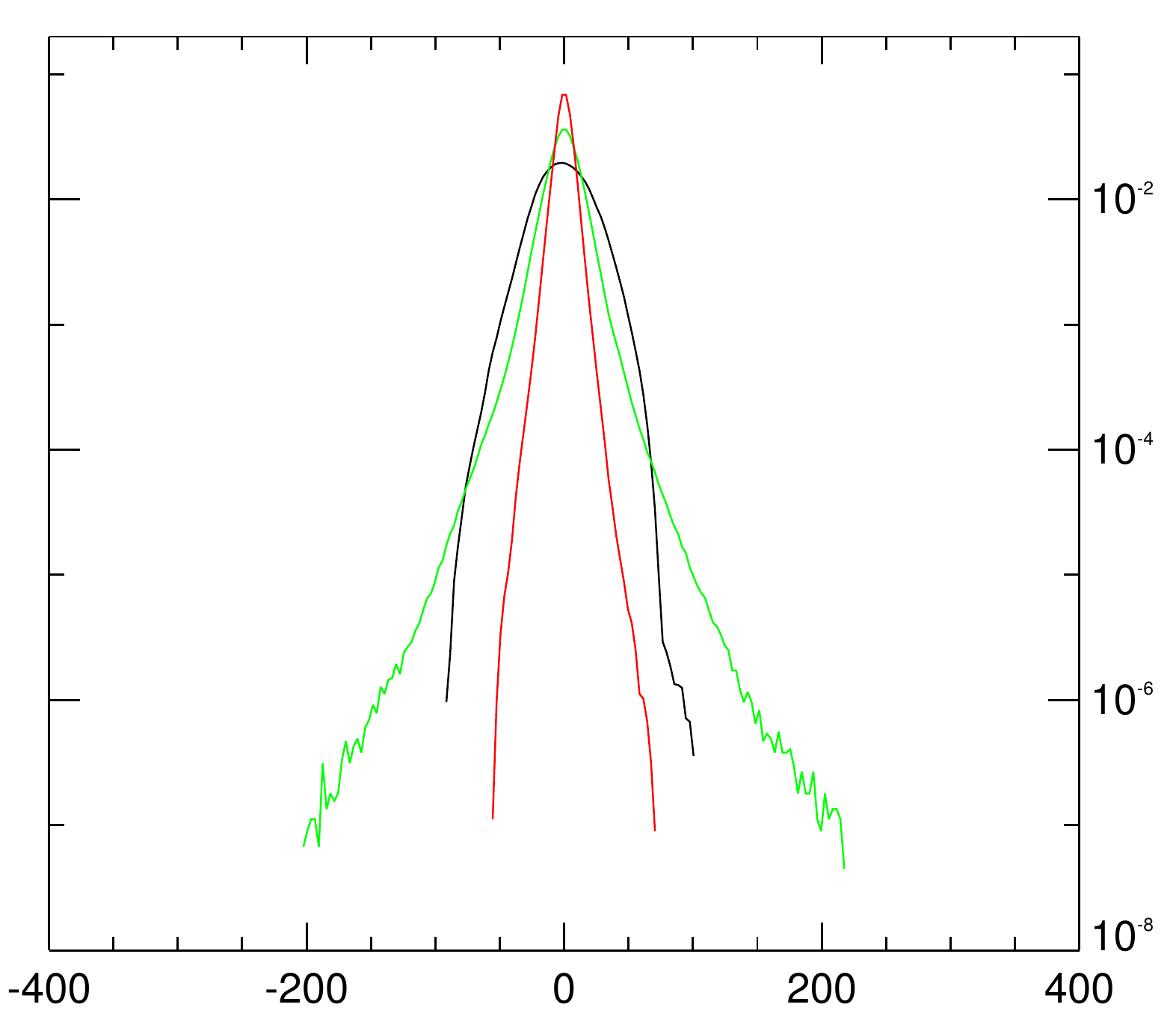}
\put(-180,165){{\large $\delta<0.5$}}
\put(-130,-5){{$\epsilon_{v,x}$ [s$^{-1}$km]}}
\put(-95,165){{\large $r_{\rm S}$=10 {\Muns}}}
\end{tabular}
\caption{\label{fig:statserr} Statistics of the errors in the reconstructed velocity for the $x$-component (green: LIN;  black: LOG-LPT, red:2LPT)  with different smoothing scales $r_{\rm S}=$ 5 (left panels) and 10 (right panels) {\Muns}. Upper panels in the entire $\delta$ range, middle panels: for the range $\delta>2$, lower panels: for the range $\delta<0.5$.}
\end{figure*}

We have calculated each component of the peculiar velocity field ($v_x$,
$v_y$,$v_z$) from the inferred velocity divergence assuming $\nabla\times\theta
= 0$.  In this case, the Fourier transform of the velocity along a direction is given by
$\hat{v} = \mbi k \cdot \hat{\theta}$ with $\mbi k$ being the k-vector. The
approximation that the velocity field is irrotational is actually a good one
for the scales and redshifts we consider here. In fact, the curl of the velocity
fields is on average more than 25 times smaller than its divergence in the z=0
output of the Millennium simulation on scales of 3 {\Muns} and even smaller
on larger scales (see Fig.~\ref{fig:sdev}).

In Fig.~\ref{fig:veldiff} we compare the velocity field along the x-axis in the
simulation and in our method (the two other axis show essentially the same
features).  The contours show the variation of number of cells (the darker
colours represent higher numbers).  By comparing both columns, it is clear that
linear theory presents biases for large speeds, which are removed by our 2LPT
approach, on both 5 and 10 {\Muns}. We can see that the distribution gets
sharper with 2LPT showing a clear decrease in the number of outliers.  

We quantified the uncertainty in the recovered velocity in
Fig.~\ref{fig:statserr}.  The x-axis corresponds to  the errors in the
reconstruction, defined as $\epsilon_{v, x}\equiv v_x^{\rm rec} -v_x$.  As in
the previous plot, we show only results for the x-axis since the other three
Cartesian coordinates provide consistent results.

We find that the errors in our method are closely Gaussian distributed -- the
skewness and kurtosis are dramatically smaller than for linear theory.  This
property is very important when applying the method to real data, since the
observational uncertainties can be added to the model uncertainties within a
Gaussian likelihood without the need of introducing complex error models.  The
typical errors are also significantly reduced with smaller standard deviations.
The errors in the reconstruction using linear theory have very long tails. Such
outliers are not present in our method (see Fig.~\ref{fig:statserr}).  

As we have discussed along this paper, the larger improvement of our method
concerns velocities in high density regions. For regions with $\delta>1$
and $\delta>2$ we find significant differences between linear and 2LPT. At 10
{\Muns} and cells with $\delta>2$ the 1 sigma errors with
linear theory are about 70 {\kms} and are reduced with 2LPT to $\sim$13 {\kms}, i.e. 
81\% smaller. The corresponding 2 sigma confidence intervals are about 160  and
28 {\kms}, respectively, i.e. an error reduction of about 83\%. One can see that
the  2 sigma confidence intervals are about double as large as the 1 sigma
confidence intervals for the 2LPT estimation.  However, this is not the case
for the linear estimates as these are not Gaussian distributed.

\section{Conclusions and discussion}

In this paper we presented an improved method to reconstruct the peculiar
velocity field from the density field. It builds from 2nd order Lagrangian
perturbation theory and the realisation that the density field can be split
into a linear plus a nonlinear term.  The latter is the key concept, which
enables the application of Lagrangian perturbation theory to an evolved field
in Eulerian coordinates. This in turn, creates an approach that is nonlinear 
and nonlocal by including the information of the gravitational tidal field 
tensor.

We have shown that this approach is efficient and accurate over the dynamical
range probed by the Millennium simulation. When the reconstruction is carried
out on 5 {\Muns}, each component of the velocity field can be recovered to
an accuracy of about 10 {\kms}. On 10 {\Muns} this figure is reduced to
about 7 {\kms}. If we consider high density regions, the typical uncertainty
is of 13  {\kms}, which improves dramatic over linear perturbation theory; typical
errors are 81\% smaller. An
accurate description of the velocity divergence, both in terms of its PDF and
on a point-by-point basis, is also achieved.  In addition, we have shown that
the 1- and 2-point statistics of the scaled velocity divergence are extremely well recovered,
being almost indistinguishable from the true ones. Contrarily, linear theory
dramatically over-estimates the velocity divergence.  This especially on the
mildly nonlinear scale of 5 {\Muns} where our method shows more clearly
its advantages.  Finally, we highlight that our method does not require calibration
nor free parameters to predict the velocity divergence field. 

There are a number of simplifications and assumptions that we have adopted
throughout our paper. First, our analyses focused on the peculiar velocity
averaged over a volume.  Another aspect is that we have neglected the
rotational component of the velocity field. This however, does not seem to be
relevant at the scales we have investigated (larger than 5 {\Muns}). Another 
simplification is performing our comparison assuming that there is an unbiased
estimation of the underlying real-space density field. But, of course, densities
measured in a survey are in redshift-space. The transformation can be done,
but not trivially. One alternative is to apply the  Gibbs-sampling method
suggested by \citet[][]{kitaura,kitaura_lyman} to correct for redshift-space distortions.  In
this, the Gaussian distribution of errors in our method is highly convenient,
since it permits to model the uncertainties in the peculiar velocity field
including observational errors in a straightforward way. 
{\color{black}   One should consider also classical iterative approaches pioneered by \citet[][]{1991ApJ...372..380Y} based on linear theory. In a similar way they have been applied with different approximations by different groups \citep[see][]{1997MNRAS.285..793C,NB00,2000MNRAS.318..681M,BEN02,2005ApJ...635L.113M,LMCTBS08,wang_sdss}.  Here it is crucial to have an accurate description relating the peculiar velocity field to the density field (or galaxy distribution) which subject of study is central in the work presented here. We expect that the improved relation found in this work with respect to linear Eulerian or linear Lagrangian perturbation theory yields better estimates of the peculiar velocity field.
 Further studies need to be done to test the performance including redshift-space distortions. }

We would like to note that our comparison and the uncertainties quoted here,
were based on the present-day output of the Millennium Run. Such simulation was
carried out with a value for $\sigma_8$ about 10\% higher than the current best
estimations \citep[see][for a method to correct for this]{2010MNRAS.405..143A}. Therefore,
our uncertainties should be regarded as an upper limit of the reachable
uncertainties for a hypothetical spectroscopic survey, which should contain a
less nonlinear underlying dark matter distribution.

{\color{black} We finalise this paper by emphasising that  the method  presented here can potentially be used in many different applications, and should be further developed and tested to perform
 bias studies combining galaxy redshift surveys with measured
peculiar velocities, Baryon Acoustic Oscillation reconstructions, determination
of the growth factor, to estimates of the initial conditions of the Universe.}

\section*{Acknowledgements}

FSK and REA thank Simon White for encouraging conversations. The authors appreciate Francisco Prada's comments on the manuscript. We are indebted to the  German Astrophysical Virtual Observatory (GAVO) and the MPA facilities  for providing us the Millennium Run simulation data.  The work of REA was supported by
Advanced Grant 246797 "GALFORMOD" from the European Research Council.
YH has been partially supported by the ISF (13/08). 

{\small
\bibliographystyle{mn2e}
\bibliography{lit}
}

\appendix

\section{Third order Lagrangian perturbation theory (3LPT)}
\label{sec:app1}

\begin{figure*}
\begin{tabular}{cc}
\includegraphics[width=8.cm]{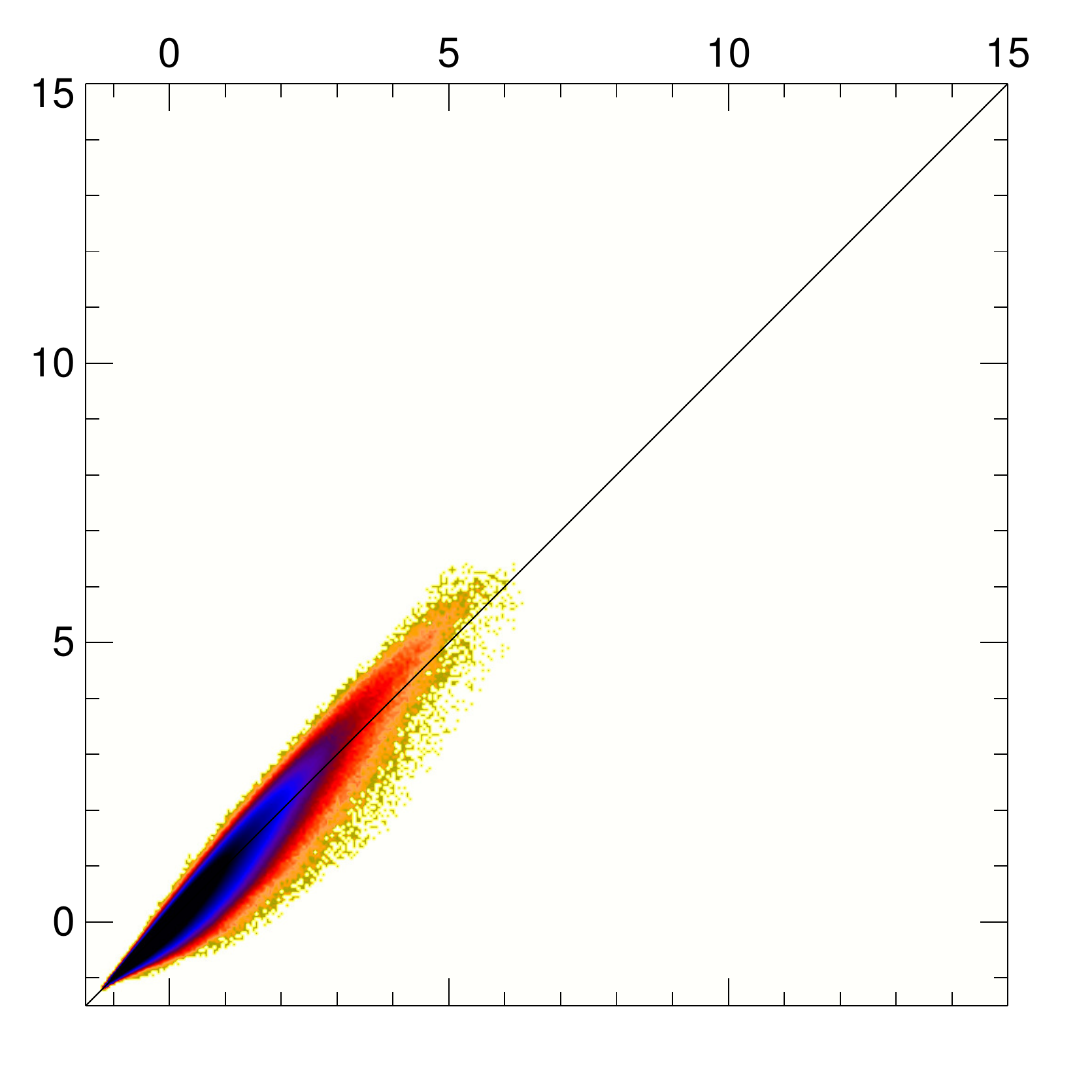}
\put(-100,70){{\large $r_{\rm S}$=5 {\Muns}}}
\put(-100,50){{\large LOG-3LPT}}
\put(-230,110){\rotatebox[]{90}{{$\theta^{\rm rec}_{\rm LOG-3LPT}$}}}
\hspace{-1.0cm}
\includegraphics[width=8.cm]{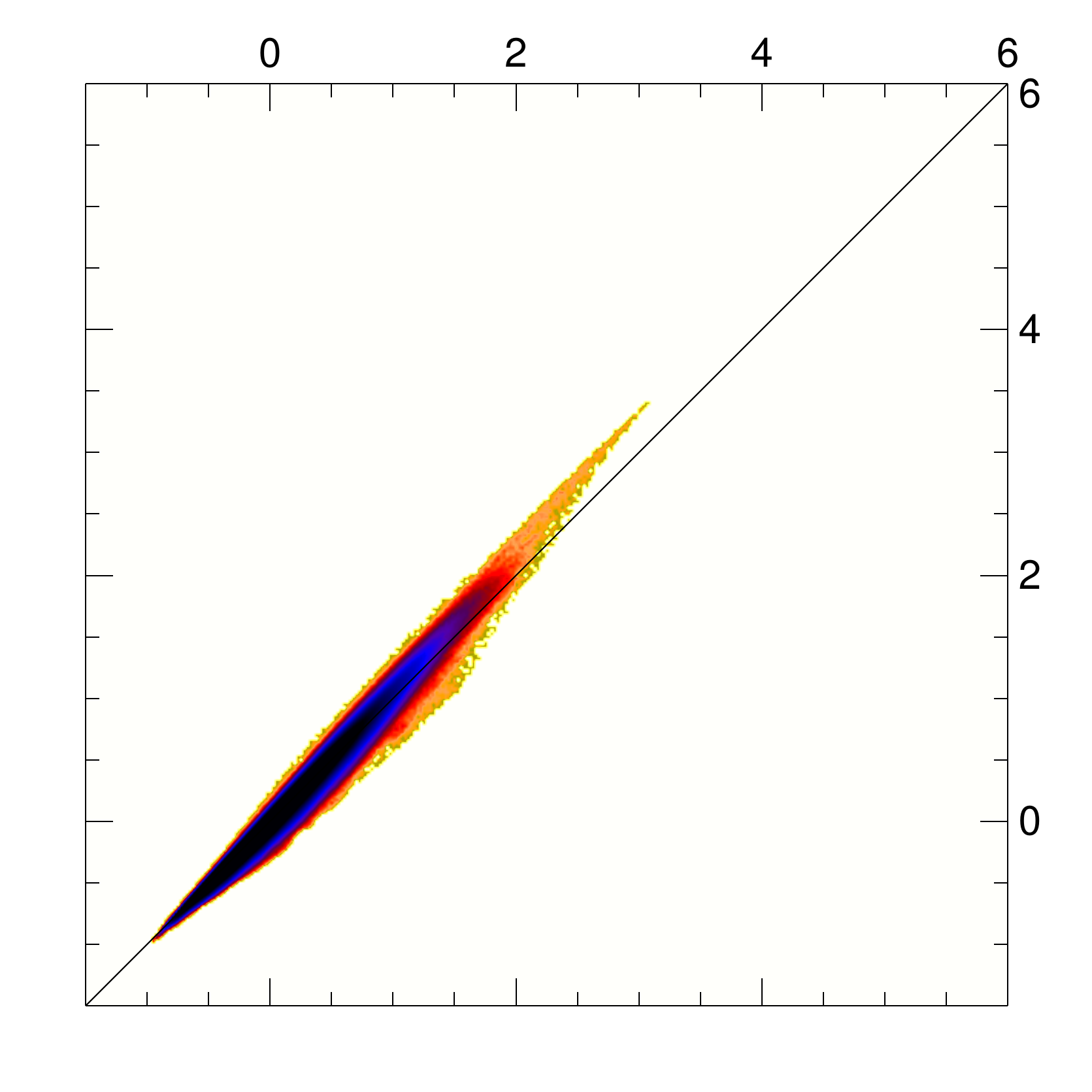}
\put(-100,70){{\large $r_{\rm S}$=10 {\Muns}}}
\put(-100,50){{\large LOG-3LPT}}
\vspace{-1.0cm}
\\
\includegraphics[width=8.cm]{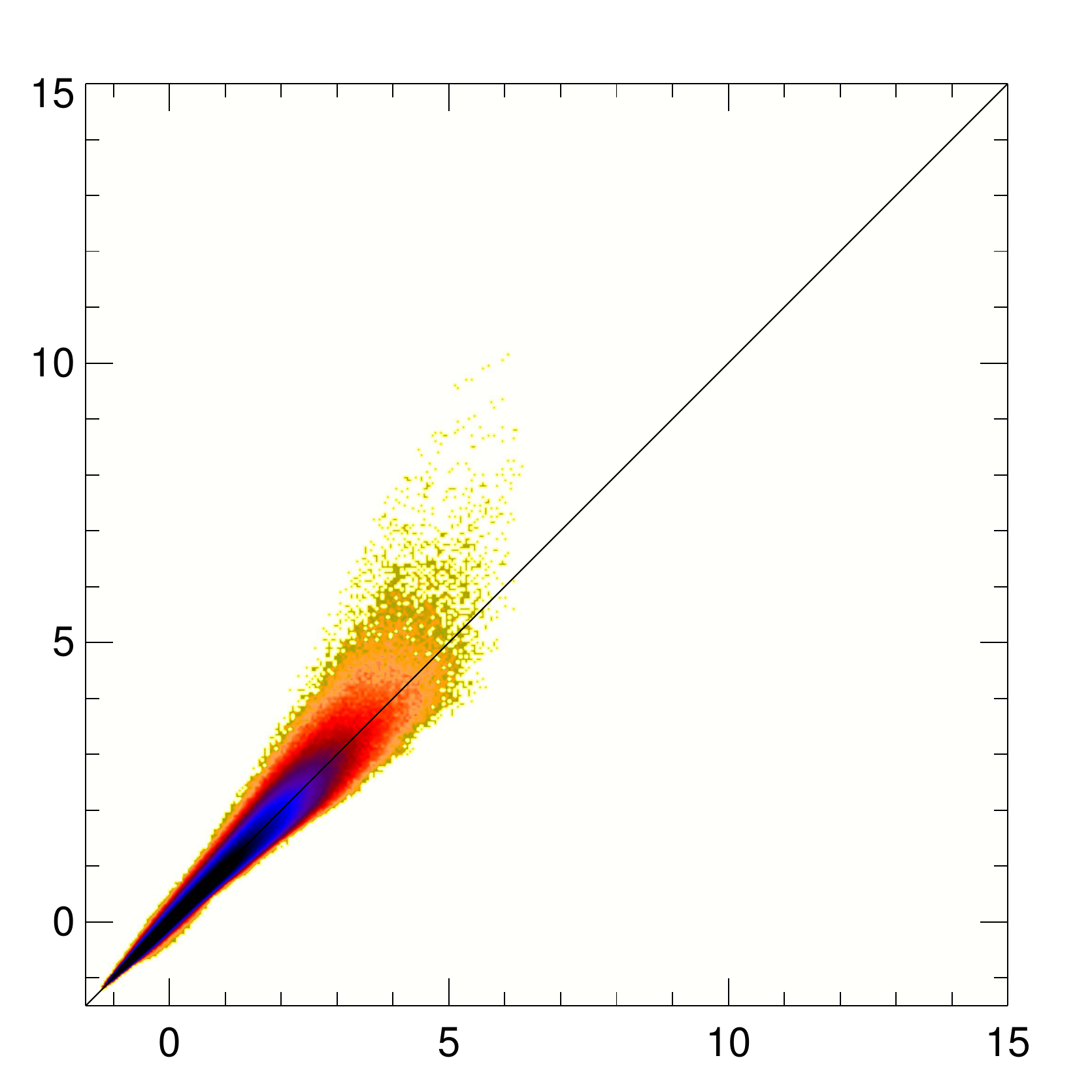}
\put(-230,110){\rotatebox[]{90}{{$\theta^{\rm rec}_{\rm 3LPT}$}}}
\put(-100,70){{\large $r_{\rm S}$=5 {\Muns}}}
\put(-100,50){{\large 3LPT}}
\put(-115,-5){{$\theta^{\rm Nbody}$}}
\hspace{-1.0cm}
\includegraphics[width=8.cm]{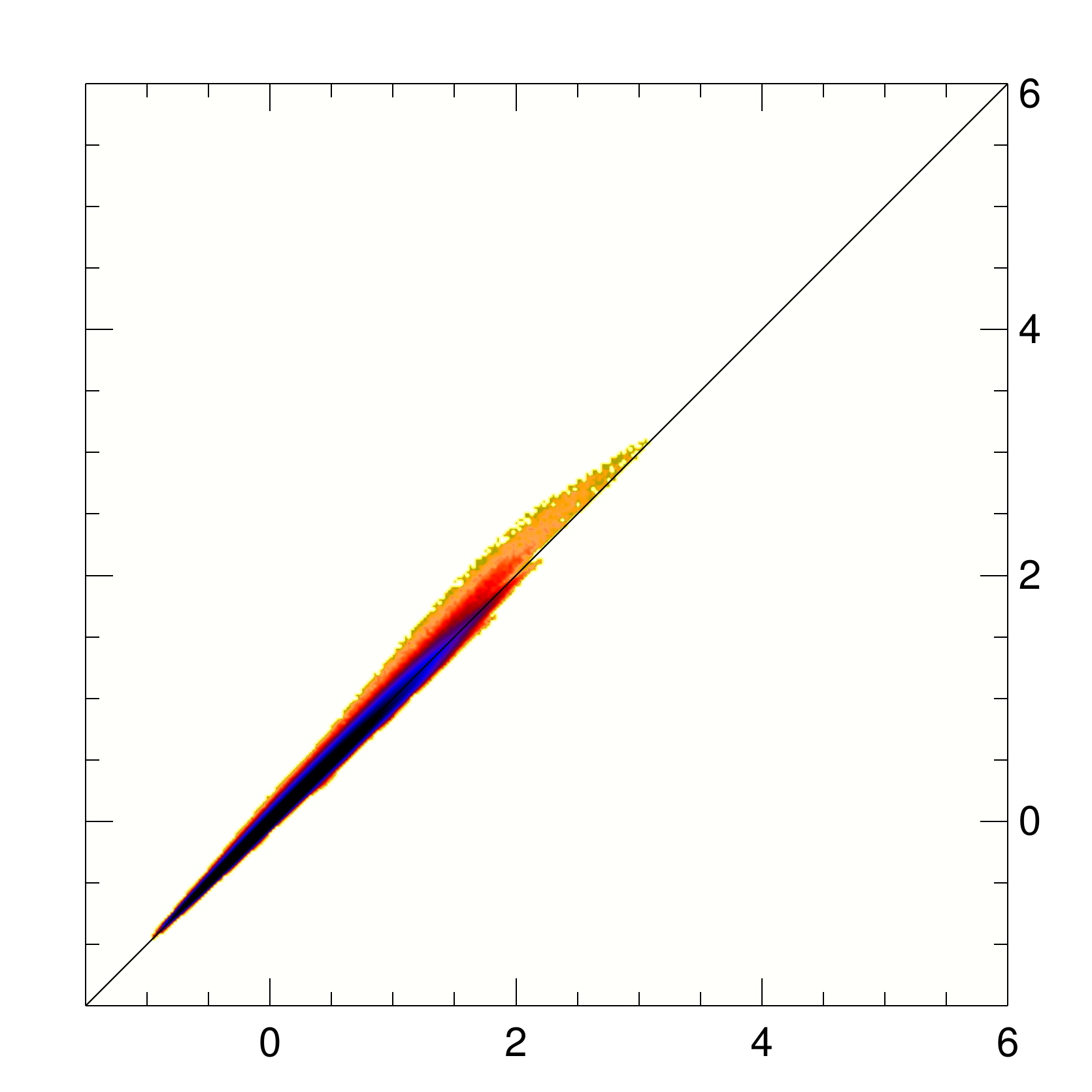}
\put(-100,70){{\large $r_{\rm S}$=10 {\Muns}}}
\put(-100,50){{\large 3LPT}}
\put(-115,-5){{$\theta^{\rm Nbody}$}}
\vspace{0.cm}
\end{tabular}
\caption{\label{fig:thetacorr3LPT} 
Cell-to-cell correlation between the {\it true} and the reconstructed scaled
velocity divergence using 3LPT. Left panels on scales of 5 {\Muns} and
right panels on scales of 10 {\Muns}. Upper panels: LOG-3LPT (based on the
truncated 3LPT expansion computing Eq.~\ref{eq:3lptthetasplit1} with the local
model for the linear component), lower panels: 3LPT (based on  the truncated
3LPT expansion computing Eq.~\ref{eq:3lptthetasplit} with the non-local model
for the linear component).} 
\end{figure*}

For the sake of completeness we investigate here third order LPT.  Following
\citet[][]{1994MNRAS.267..811B} and \citet[][]{1995MNRAS.276..115C} one can
write the displacement field to third order as

 \ba
\label{eq:3lpt} 
 \mbi \Psi & =&  - D_1\nabla\phi^{(1)} + D_2\nabla\phi^{(2)}\nonumber\\
&&+D_{3{\rm a }}\nabla\phi^{(3)}_{{\rm a }}+D_{3{\rm b}}\nabla\phi^{(3)}_{{\rm b}}+D_{3{\rm c}}\nabla\times\mbi A^{(3)}
\ea

\noindent where \{$D_{3{\rm a }},D_{3{\rm b}},D_{3{\rm c}}$\} are the 3rd
order growth factors corresponding to the gradient of two scalar potentials
($\phi^{(3)}_{{\rm a }}$,$\phi^{(3)}_{{\rm b }}$) and the curl of a vector
potential ($\mbi A^{(3)}$). Particular expressions for the irrotational 3rd
order growth factors ($D_{3{\rm a }},D_{3{\rm b}}$) can be found in
\citet[][]{bouchet1995}, the growth factor of the rotational term ($D_{3{\rm
c}}$) is given in \citet[][]{1995MNRAS.276..115C}.

We assume that the fields are curl-free on scales $\gsim$ 5 {\Muns}
\citep[see][and  the comparison between the velocity divergence and the curl in
the Millennium Run in Fig.~\ref{fig:sdev}]{bouchet1995}.  We therefore consider
only the scalar potential terms $\phi^{(3)}_{{\rm a }}$ and $\phi^{(3)}_{{\rm b
}}$.

The first term is cubic in the linear potential

\be
\delta^{(3)}_{{\rm a}}\equiv\mu^{(3)}(\phi^{(1)})=\det\left(\phi^{(1)}_{,ij}\right)\,,
\ee


\noindent and the second term is the interaction term between the first- and the
second-order potentials:

\be
\label{eq:3lptdelta2}
 \delta^{(3)}_{{\rm b}}\equiv\mu^{(2)}(\phi^{(1)},\phi^{(2)})=\frac{1}{2}\sum_{i\ne j} 
    \Big( \phi^{(2)}_{,ii}\phi^{(1)}_{,jj}-
    \phi^{(2)}_{,ij}\phi^{(1)}_{,ji}\Big)\,,
\ee

\noindent \citep[see][]{1994MNRAS.267..811B,bouchet1995,1995MNRAS.276..115C}.  In order
to ensure that the term $\delta^{(3)}_{{\rm b}}$ is curl-free one has to
introduce some proper weights in the expression \ref{eq:3lptdelta2}
\citep[see][]{1994MNRAS.267..811B,1995MNRAS.276..115C}.

From the displacement field one can derive the expression for the velocity field

 \ba
\label{eq:3lptv} 
 \lefteqn{\mbi v =  - D_1f_1H\nabla\phi^{(1)} + D_2f_2H\nabla\phi^{(2)}}\\
&&\hspace{-.0cm}+D_{3{\rm a }}f_{3{\rm a }}H\nabla\phi^{(3)}_{{\rm a }}+D_{3{\rm b}}f_{3{\rm b}}H\nabla\phi^{(3)}_{{\rm b}}\nonumber\,,
\ea

\noindent with  $f_i = {\rm d}\ln D_i/{\rm d}\ln a$ \citep[particular
expressions for $f_{3{\rm a }}$ and $f_{3{\rm b}}$ can be found
in][]{bouchet1995}.  As we can see from Eq.~\ref{eq:3lptv} one can construct
all components from the linear potential $\phi^{(1)}$.

We consider here two models for the linear potential. The first model relies on
the local lognormal estimate  (see \S \ref{sec:method}) from which the scaled
velocity divergence can be calculated in the following way to 3rd order LPT

\ba
\label{eq:3lptthetasplit1}
\lefteqn{\theta=D_1\delta^{(1)}-\frac{f_2}{f_1}D_2\mu^{(2)}(\phi^{(1)})}\nonumber\\
&&\hspace{-0.cm}-\frac{f_{3{\rm a }}}{f_1}D_{3{\rm a }}\mu^{(3)}(\phi^{(1)})-\frac{f_{3{\rm b}}}{f_1}D_{3{\rm b}}\mu^{(2)}(\phi^{(1)},\phi^{(2)})\,.
\ea

The second model yields a non-local estimate of $\phi^{(1)}$ to 3rd order given
by 

\be
\label{eq:3lptsplit}
\delta^{\rm L}=\delta-\delta^{\rm NL}\,
\ee

\noindent with  $\delta^{\rm NL}=-D_2\mu^{(2)}(\phi^{(1)})-D_{3{\rm a
}}\mu^{(3)}(\phi^{(1)})-D_{3{\rm
b}}\mu^{(2)}(\phi^{(1)},\phi^{(2)})+\mu^{(2)}(\Theta)+\mu^{(3)}(\Theta)$.  Note
that the potential $\Theta$ is also different and is determined by
Eq.~\ref{eq:3lpt} recalling the relation to the displacement field $\mbi
\Psi=-\nabla \Theta$. 

Using the latter expression one can write the $\theta$-$\delta$ relation to 3rd
order in LPT as

\ba
\label{eq:3lptthetasplit}
\lefteqn{\theta=\delta-\left(\frac{f_2}{f_1}-1\right)D_2\mu^{(2)}(\phi^{(1)})}\nonumber\\
&&\hspace{-0.5cm}-\left(\frac{f_{3{\rm a }}}{f_1}-1\right)D_{3{\rm a }}\mu^{(3)}(\phi^{(1)})-\left(\frac{f_{3{\rm b}}}{f_1}-1\right)D_{3{\rm b}}\mu^{(2)}(\phi^{(1)},\phi^{(2)})\nonumber\\
&&\hspace{-0.5cm}-\mu^{(2)}(\Theta)-\mu^{(3)}(\Theta)\,.
\ea

Note that Eq.~\ref{eq:3lptsplit} should be solved iteratively as we do in the
2LPT case \citep[see \S \ref{sec:method} and][]{kitaura_gauss}. Nevertheless to
find a fast solution we plug-in the lognormal model for $\phi^{(1)}$ into
Eq.~\ref{eq:3lptthetasplit1} yielding a non-local estimate of the linear field
in one iteration. To minimize the deviation between LPT and the full nonlinear
evolution we have additionally smoothed the density $\delta$ in
Eq.~\ref{eq:3lptthetasplit} with a Gaussian kernel of 2 {\Muns} radius. 
 
We do not find  an improvement in the determination of the velocity divergence
with respect to 2LPT including both curl-free terms using any of the estimates
of the linear component.  This could be due to an inaccurate determination of
the interaction term $\phi^{(3)}_{{\rm b}}$, since uncertainties in the linear
component $\phi^{(1)}$ propagate more dramatically than in terms which depend
only on the initial conditions ($\phi^{(2)},\phi^{(3)}_{{\rm a}}$).

One may neglect the interaction term $\delta^{(3)}_{{\rm b}}$ and consider only
the cubic term $\delta^{(3)}_{{\rm a}}$ to 3rd order. Such a truncated 3LPT
model includes the {\it main body} of the perturbation sequence with the rest
of the sequence being made up of interaction terms
\citep[][]{1994MNRAS.267..811B}.

Our calculations show in this case better results than 2LPT for low values of
the velocity divergence (see Fig.~\ref{fig:thetacorr3LPT}). However, for large
values of $\nabla\cdot \mbi v$ we obtain larger dispersions which could be also
due to numerical errors in the estimate of the linear density component
$\delta^{\rm L}$.  The errors in the velocity estimation are only moderately
reduced with respect to the 2LPT case (see \S \ref{sec:vres}).

\end{document}